\documentclass[reprint,twocolumn,aps,prc,a4paper,superscriptaddress,showpacs,preprintnumbers,amsmath,amssymb]{revtex4}
\usepackage{graphicx}
\usepackage{dcolumn}
\usepackage{bm}
\usepackage{CJK}
\usepackage{float}
\usepackage{amssymb} 
\usepackage{amsmath}
 
\begin{document}

\title{Toward precision mass measurements of neutron-rich nuclei relevant to $r$-process nucleosynthesis}
\author{B. H. Sun}\thanks{e-mail: bhsun@buaa.edu.cn}
 \affiliation{School of Physics and Nuclear Energy Engineering, Beihang University, Beijing 100191, China}
 \affiliation{International Research Center for Nuclei and Particles in the Cosmos, Beijing 100191, China}
 \author{Yu. A. Litvinov} 
\affiliation{Key Laboratory of High Precision Nuclear Spectroscopy and Center for Nuclear Matter Science, Institute
of Modern Physics, Chinese Academy of Sciences, Lanzhou 730000,
People's Republic of China}
\affiliation{GSI Helmholtzzentrum f\"{u}r Schwerionenforschung,
Planckstra{\ss}e 1, 64291 Darmstadt, Germany} 
\author{I. Tanihata}
 \affiliation{School of Physics and Nuclear Energy Engineering, Beihang University, Beijing 100191, China}
 \affiliation{International Research Center for Nuclei and Particles in the Cosmos, Beijing 100191, China} 
 \author{Y.~H.~Zhang}
 \affiliation{Key Laboratory of High Precision Nuclear Spectroscopy and Center for Nuclear Matter Science, Institute of Modern Physics, Chinese Academy of
Sciences, Lanzhou 730000, People's Republic of China} 
 
\date{\today}
\begin{abstract}

The open question of where, when, and
how the heavy elements beyond iron enrich our Universe has triggered
a new era in nuclear physics studies. Of all the relevant nuclear
physics inputs, the mass of very neutron-rich nuclides is a key
quantity for revealing the origin of heavy elements beyond iron.
Although the precise determination of this property is a great
challenge, enormous progress has been made in recent decades, and it
has contributed significantly to both nuclear structure and
astrophysical nucleosynthesis studies. In this review, we first
survey our present knowledge of the nuclear mass surface,
emphasizing the importance of nuclear mass precision in $r$-process
calculations. We then discuss recent progress in various methods of
nuclear mass measurement with a few selected examples. For each
method, we focus on recent breakthroughs and discuss possible ways
of improving the weighing of $r$-process nuclides.

\tableofcontents


\end{abstract}
\maketitle
\date{today}

\section{Introduction}\label{sec:introduction}

\begin{figure*}[th]
   \centering
   \includegraphics[width=15cm]{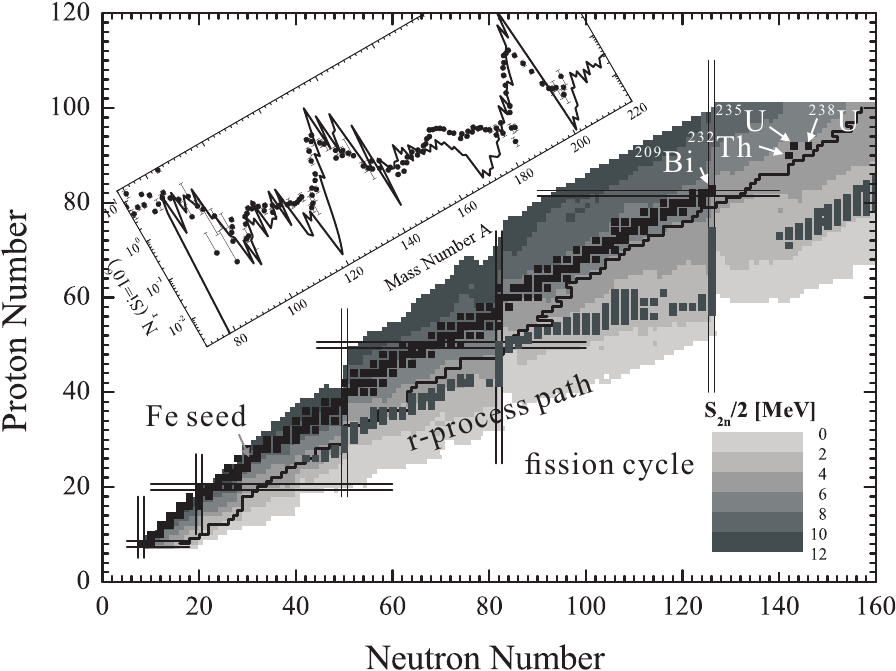}
   \vspace{-10pt} \caption{Features of the $r$-process calculated using
   the RMF-BCS mass table. Black squares denote $\beta$-stable nuclei,
   and magic proton and neutron numbers are indicated by pairs of
   parallel lines. The region in the main graph shows the calculated
   average one neutron separation energy ($S_{2n}/2$). The solid line
   denotes the border of nuclides with known masses in the neutron-rich
   side. The dark grey squares show the $r$-process path when using the
   RMF mass predictions and the FRDM half-lives. The observed and
   calculated solar $r$-process abundance curves are plotted versus the
   mass number $A$ in the inset, whose x-axis is curved slightly to
   follow the $r$-process path. Adapted from Ref.~\cite{Sun2008Chin.Phys.Lett.2429}.} \label{fig1}
\end{figure*}

Elements and their distribution in nature carry the signatures of astrophysical processes that occurred long before our sun was born. Indeed, the solar abundance distribution played a crucial role in studies on the origin of the elements. To account for this ``universal'' elemental distribution, various nucleosynthesis processes have been proposed.
Among them, the rapid neutron-capture process ($r$-process) is generally accepted to be one of the major mechanisms producing the stable neutron-rich nuclides (and some long-lived radioactive ones, such as $^{99}$Tc) from about Ga, element 31, up to long-lived radioactive uranium, element 92~\cite{BBFH,1957-Cameron,Cowan1991PRp,Qian2007,Arnould2007}.
These isotopes are observed in stars of different metallicities as well as in the solar system (for a review, see Ref.~\cite{2008-Sneden-ARAA-abundance}). 

The $r$-process is somehow unique in the sense that it is the only known process that
can produce elements heavier than Bi and reach thorium and uranium, and furthermore the only known process that can possibly synthesize superheavy elements in nature. 
A widely accepted picture is that the $r$-process occurs under conditions in which neutrons with densities of $10^{20}-10^{30}$ cm$^{-3}$ are captured on a very fast timescale of milliseconds.  
It starts with the reactions from hydrogen and is terminated when the heaviest nuclei created in the process become unstable to spontaneous fission. The abundance flow runs up along the contour lines near the neutron drip line, and as it does so, highly unstable neutron-rich nuclei are created.  A schematic view of the $r$-process path is shown in Fig.~\ref{fig1}. Accordingly, several thousands of nuclei lying between the $\beta$-stability line and the neutron drip line are involved in the matter flow.
Understanding the $r$-process, therefore, requires knowledge of properties such as the masses, $\beta$-decay lifetimes, and neutron-capture cross sections for a few thousands of neutron-rich nuclei far from stability~\cite{Cowan1991PRp,Qian2007,Arnould2007,Sun2008Chin.Phys.Lett.2429}.

In fact, already in the 1950s, the pioneers in the field were aware of the close link  between elemental abundance peaks (at mass numbers of around 85, 130, and 195) and
the location of closed neutron shells (at 50, 82, and 126)~\cite{BBFH,1957-Cameron}. 
Since then, joint work consisting of astronomical observations, nuclear experiments and advances in theory, and nucleosynthesis modeling have significantly accelerated the development
and deployment of new solutions and projects worldwide for $r$-process-motivated studies.
Taking the nuclear mass, the key nuclear physics input for $r$-process simulations and
the fundamental properties of atomic nuclides, as an example, there are indeed remarkable
efforts in theoretical and experimental nuclear physics to reliably determine this key observable~\cite{Lunney2003Rev.Mod.Phys.1021,Blaum2006PhysRep425,Franzke2008MSR27}.
These advances have further provided solid support to $r$-process simulations.
At the risk of being partial and incomplete, we
showcase the progress in mass determination of neutron-rich nuclides relevant to the $r$-process
by considering a few selected topics, with the hope that they will provide a good sample
of the much fuller tapestry of the advancement of this very exciting field. We do not intend to describe in detail each mass measurement technique, 
but focus only on the principles, the achievements realized by applying each technique, their main limitations, and possible ways to overcome those limitations.

The paper is organized as follows. In Sec.~\ref{sec:precision}, we first stress the importance of the precision of nuclear masses in modeling the $r$-process
and then summarize how well we know the nuclear masses globally.  
Particular emphasis is then placed on the modern experimental methods for weighing the masses of $r$-process nuclei in Sec.~\ref{sec:progress}.
Here, rather than a detailed explanation of the principles of various experimental techniques, we discuss how different methods can be improved 
toward mass measurements of nuclei relevant to $r$-process studies. In Sec.~\ref{sec:remark}, we discuss the difficulties of measuring n-rich nuclei, the necessity 
of having different methods as an independent cross-check, and comparisons of the methods based on the challenges of applying them.  Finally, we present the conclusion in Sec.~\ref{sec:conclusion}. Throughout the paper,
we also mention briefly the opportunities for future nuclear mass measurements 
in China.

\section{Nuclear mass and precision}
\label{sec:precision}

Nuclear masses probably have the most decisive influence on the operation of the $r$-process.
They determine the position of the neutron drip line,
the neutron separation energies (i.e., the $Q$ values of neutron captures),
and the $Q$ values for beta decay  and nuclear reactions.
The precision of the nuclear mass that we need for $r$-process studies has been stressed in many papers~(e.g., \cite{Lunney2003RMP75,Blaum2006PhysRep425}). 
Generally, a precision of $\delta m/m$
in the order of around 10$^{-6}$ is needed, whereas 10$^{-7}$ or even better is necessary for specific cases such as
waiting-point nuclei in the $r$-process path. Here we illustrate the importance of the precision
of nuclear masses in $r$-process studies by a specific example.

In the waiting-point approach~\cite{Cowan1991PRp,Sun2008Phys.Rev.C025806, Niu2009Phys.Rev.C065806,Xu2013PhysRevC.87.015805}, the abundance ratio between two neighboring isotopes is given by the Saha equation:
  \begin{eqnarray}     \label{eq:equilibrium}
    \frac{Y(Z,A+1)}{Y(Z,A)} &=& n_n\left(\frac{2\pi\hbar^2}{m_u k T}\right)^{\frac{3}{2}}
    \frac{G(Z,A+1)}{2G(Z,A)}\cr
          & & \left(\frac{A+1}{A}\right)^{3/2} \exp\left[\frac{S_n(Z,A+1)}{\kappa T}\right]\;,
  \end{eqnarray}
where $\hbar$ is the Planck constant, $m_u$ is the atomic mass unit, and $k$ is the Boltzmann
constant. $(Z,A)$ indicates a nucleus with proton number $Z$ and mass number $A$.
 $Y$, $G$, and $S_n$ denote the number abundance, partition function, and neutron
separation energy of the appropriate nucleus, respectively. The astrophysical conditions in the waiting-point approximation 
are characterized by the temperature $T$ and neutron density $n_n$.
For a specific isotopic chain, the corresponding waiting-point nucleus has the largest abundance and
is determined by the partition functions and neutron separation energies of the relevant
nuclei for fixed $T$ and $n_n$.

\begin{figure}
\centerline{
\includegraphics[angle=0, width=0.42\textwidth]{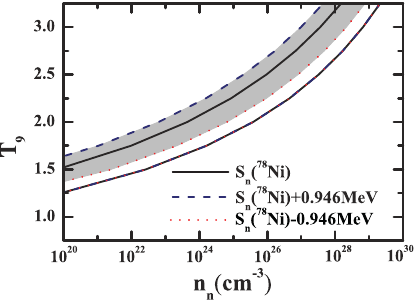}
} \caption{(Color online) Effects of the uncertainty in the neutron separation energy $S_n$ for $^{78}$Ni on the $T_9$-$n_n$ conditions required by the
$N=50$ waiting point nuclei $^{80}$Zn, $^{79}$Cu and $^{78}$Ni. The band between two curves of the same
kind represents the conditions using different $S_n$ of $^{78}$Ni. The shaded regions in each panel show the effects on the required conditions when the corresponding $S_n$ values are varied within the estimated uncertainties. $T_9$ is defined as $10^9$ K. }
\label{fig2}
\end{figure}

As can be seen from the exponential dependence on
the neutron separation energy in Eq.~(\ref{eq:equilibrium}), $Y(Z,A+1)/Y(Z,A)$ depends mainly on the neutron separation energy.
We ignore the small differences in the nuclear partition function and mass number and rewrite Eq.~(\ref{eq:equilibrium}) as
\begin{equation}\label{eq:equilibrium2}
\frac{Y(Z,A+1)}{Y(Z,A)}=\exp\left[\frac{S_n(Z, A+1)-S_n^0(T,n_n)}{k T}\right],
\end{equation}
where
\begin{eqnarray}\label{eq:Sn0}
S_{n}^{0}(T,n_n)&\equiv& kT\ln\left[\frac{2}{n_n}\left(\frac{m_u kT}{2\pi\hbar^2}\right)^{3/2}\right] = 2.79T_9 \cr
& &  + 0.198T_9\left[\log_{10}
     \left(\frac{10^{20}}{n_n}\right)+\frac{3}{2}\log_{10} T_9\right] \;.
\end{eqnarray}
 
In the second equality of Eq.~(\ref{eq:Sn0}), $T_9$ is $T$ in units of $10^9$~K, and $S_n$ is in mega-electron volts. Ideally, $S_n$ is the same for 
nuclei along the $r$-process path for a given neutron density $n_n$ and temperature $T_9$. 
In other words, the $r$-process proceeds along lines of constant neutron separation energies toward heavy nuclei. Higher temperature or lower neutron density will drive the $r$-process path toward the valley of stability. For typical $r$-process conditions, this corresponds to $S_n \sim 2$--4~MeV. However, the separation energy is not a smoothly changing function of the neutron number but shows large jumps, particularly close to magic neutron numbers. This is also the reason that the $r$-process path moves closer to stability where the relevant nuclei have larger $S_n$. Considering the pairing correlations, the most abundant isotope always has an even neutron number $N$.
 
Eq.~(\ref{eq:equilibrium}) makes it possible to deduce the required $T_9$--$n_n$ conditions for the $r$-process, as the required conditions
are determined mainly by the neutron separation energies of some crucial waiting points (CWPs) around neutron magic numbers 50, 82, and 126, and of those nuclei around them~\cite{Cowan1991PRp,Xu2013PhysRevC.87.015805}.
As an example, we show the importance of knowing the mass of nuclei at and beyond $N=50$ along the $r$-process path, as these nuclei represent the few cases that  experiments may be able to access in the near future. 
Fig.~\ref{fig2} shows the only conditions, characterized by $T_9$ and $n_n$, under which the $N=50$ CWP nuclides $^{78}$Ni, $^{79}$Cu, and $^{80}$Zn can be produced.
For a specific $T_9$, the values of $n_n$ between two identical lines in this figure
would allow all these CWPs to have $\geq50\%$ of the total abundance of
its isotopic chain. Furthermore, Fig.~\ref{fig2} also indicates the dramatic changes in the conditions required for $^{78}$Ni when
its neutron separation energy is varied within the estimated uncertainty of 0.946~MeV
\cite{Audi2011_Private} while the other inputs are kept the same.
Increasing the neutron separation energy of
$^{78}$Ni by 0.946~MeV raises the upper bound from
the solid curve (upper bound in Fig.~\ref{fig2}) to the dashed curve, and
decreasing this quantity by the same amount lowers it to the dotted curve.
The lower bound on the region of the required $T_9$--$n_n$ conditions in Fig.~\ref{fig2} stays the same
because it is determined by the two-neutron separation energy of $^{82}$Zn.

 \begin{figure}[h]
 \centerline{
 \includegraphics[width=0.45\textwidth]{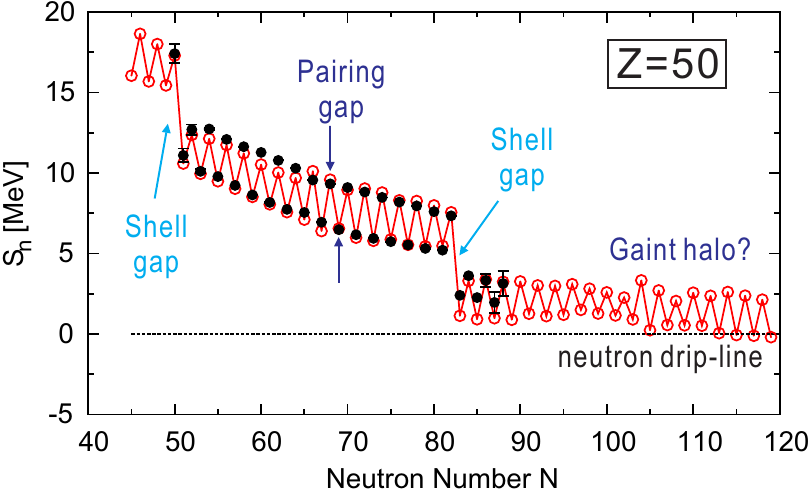}}
  \caption{Calculated (open circles) vs. experimental (filled circles) one-neutron separation energies for the Sn isotopes. Pairing gap, shell gap, neutron drip-line and giant halo are indicated. The theoretical values are from Ref.~\cite{Geng2005PTP}.}
 \label{fig3}
 \end{figure}

The significant effects of uncertainties in the neutron separation
energies on the required $T_9$--$n_n$ conditions, as shown in  Fig.~\ref{fig2},
clearly demonstrate the importance of precise mass measurements for $^{76}$Ni to $^{78}$Ni. Using this approach,
one can identify the key nuclei, including $^{76}$Ni to $^{78}$Ni, $^{82}$Zn, $^{131}$Cd, and $^{132}$Cd~\cite{Xu2013PhysRevC.87.015805}. These nuclei  at neutron shells N=50 and 82  
have the largest impact and are important candidates with high priority for precise mass measurements at rare-isotope beam (RIB) facilities.  Similarly situation may occur also at N=126. However, experimental data are still too scare in this region to allow for a precise identification of the corresponding key nuclei.

 \begin{figure*}
\includegraphics[angle=0, width=0.8\textwidth]{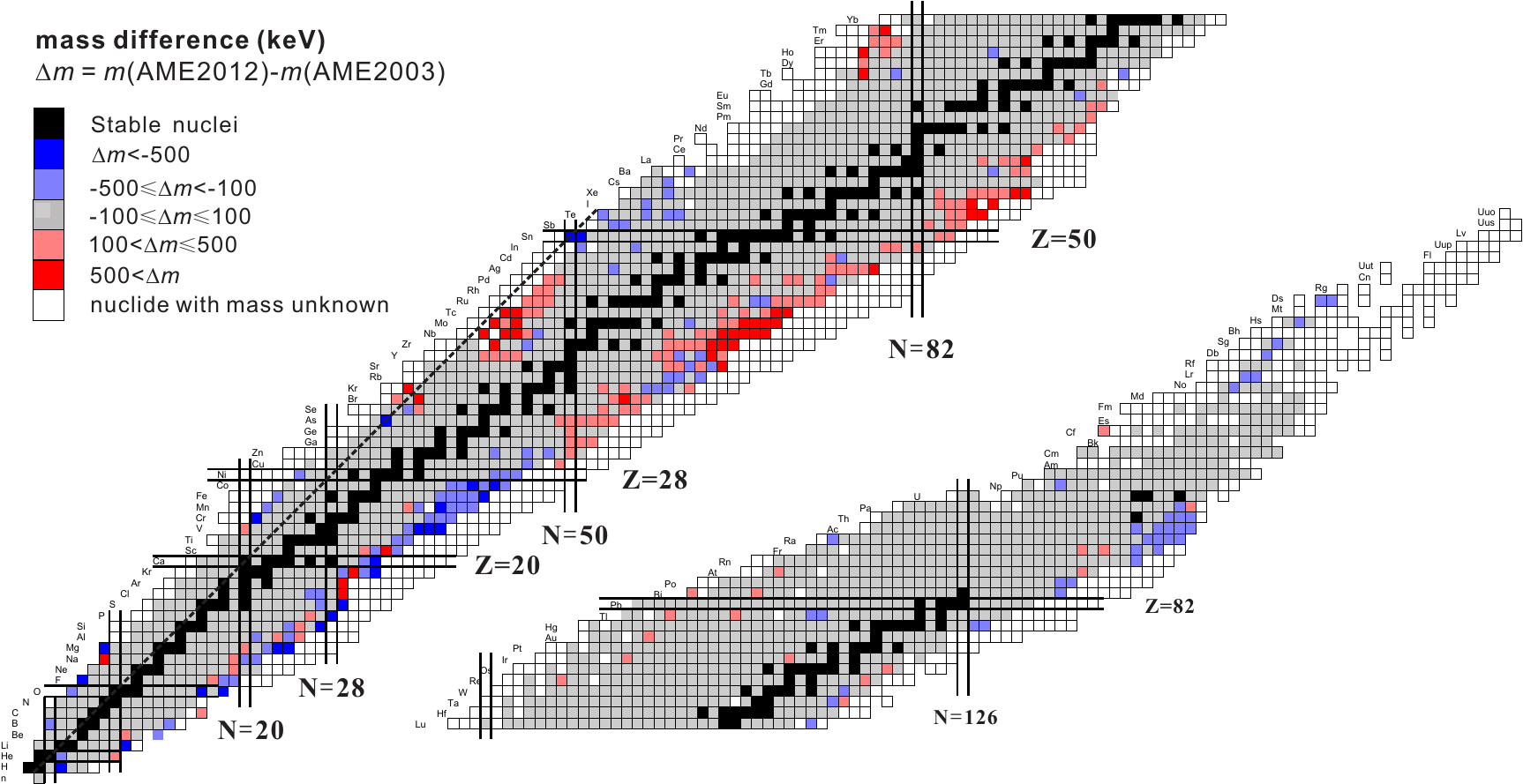}
\caption{Mass differences between two recent mass evaluation databases AME2003 and AME2012.
Experimentally mass-known nuclides (color coded squares) are represented by a color-coded cell, while
the black squares indicate stable isotopes. Magic numbers and the $N = Z$ line are
indicated by pairs of parallel lines and a dashed line, respectively. Nuclides with unknown mass are indicated by open squares.}
\label{fig-chart}
\end{figure*}

In addition to its importance in nuclear astrophysics, accurate information on nuclear masses can promote the understanding of many features of atomic nuclides.
The mass or total binding energy itself reflects all the interactions of nucleons inside a nuclide and the stability of nuclides.
Its first and second derivatives give the one/two-nucleon separation energies and accordingly enable experimental determination of drip lines,  shell gaps, and pairing gaps. A particularly interesting case
is the so-called ``giant'' halo predicted by the relativistic mean field approach~\cite{Meng1998Phys.Rev.Lett.460,Meng2002Phys.Rev.C041302,Meng2006Prog.Part.Nucl.Phys.470}. 
Although the predicted giant halo region is far from the
known experimental mass surface, this calculation opens a fantastic possibility in exotic nuclei when more neutrons are added.
A schematic view of masses in nuclear structure is shown in Fig.~\ref{fig3}. 
In the last decade, the new ultrahigh level of precision has led to a reexamination
of many contributions to nuclear properties such as pairing interactions~\cite{Litvinov2005PhysRevLett.95.042501}, shell closures~\cite{Sun2008Nucl.Phys.A812,JYFLTRAP2012PhysRevLett.109.032501}, 
the residual interaction of the last valence neutron and last valence proton~\cite{Chen2009PRL102}, and three-nucleon forces~\cite{TITAN-PhysRevLett.109.032506,MRTOF-ISOTRAP2013Nature}.

Advances in new and precise mass measurements revealed, moreover, large deviation from previous experiments when moving away from the $\beta$-stability line, as shown in  Fig.~\ref{fig-chart}.
The new mass values of neutron-rich nuclides above Ni isotopes are generally larger than those tabulated in the AME2003~\cite{AME2003}; the deviation can be up to around 2 MeV,
but is smaller between neutron shells 28 and 50.
This systematic shift in the mass surface to either larger or smaller
values can have several consequences. First, the systematically
underestimated/overestimated mass values
in the AME2003 can misguide the development of various
mass models, which are normally fitted to the experimentally known masses 
(and maybe also to other properties). The resulting
incorrect isospin dependence of nuclear masses alters
predictions regarding the basic nuclear structure, such as the evolution of shell closures, deformations,
the pairing strength, and the location of drip lines. This
feature will require more careful judgment in developing nuclear
theoretical models, especially mass-fit-driven theoretical mass models. Second, this results in inaccurate
extrapolations of AME2003, as exotic nuclear systems
are quite different in the last known neutron-rich nuclides. Modification of the extrapolated mass surface is clearly seen in the latest mass evaluation~\cite{AME2012}.

This large deviation between two mass evaluations further demonstrates the necessity of having various experimental techniques for mass measurements,
which on the one hand are complementary in most cases, and on the other hand can provide a valuable cross-check. Both aspects are essential. 
A classic example of the employment of different methods
is the storage ring and Penning trap facilities appearing worldwide.

\section{Progress in mass measurements of neutron-rich nuclides}\label{sec:progress}

In conventional mass spectrometry~\cite{Munzenberg20139}, ``simple'' magnetic and electric sector-field separators were used for mass measurements of nuclides near the beta-stability line. 
Different from this,  new techniques are currently emerging at many facilities as a premier resource for  mass measurements of atomic nuclei. 
They aim on the one hand to increase the overall efficiency and precision of mass measurements of atomic nuclei, and on the other hand to obtain access
to more unknown exotic systems.
Both goals require making the most of existing facilities and developing more effective, sensitive, and precise detection systems.
Indeed, together with the advances in the availability of intense, stable 
radioactive beams worldwide, breakthroughs in ion-beam cooling, manipulation, and detection allow
access to more exotic nuclear systems in their ground and/or isomeric states even in a single experiment.
As a result, the masses of more than 220 nuclides were determined for the first time in the last 10 years (see Fig.~\ref{fig-chart} for the current known nuclear mass surface).

\begin{figure*}
\centerline{
\includegraphics[angle=0, width=0.8\textwidth]{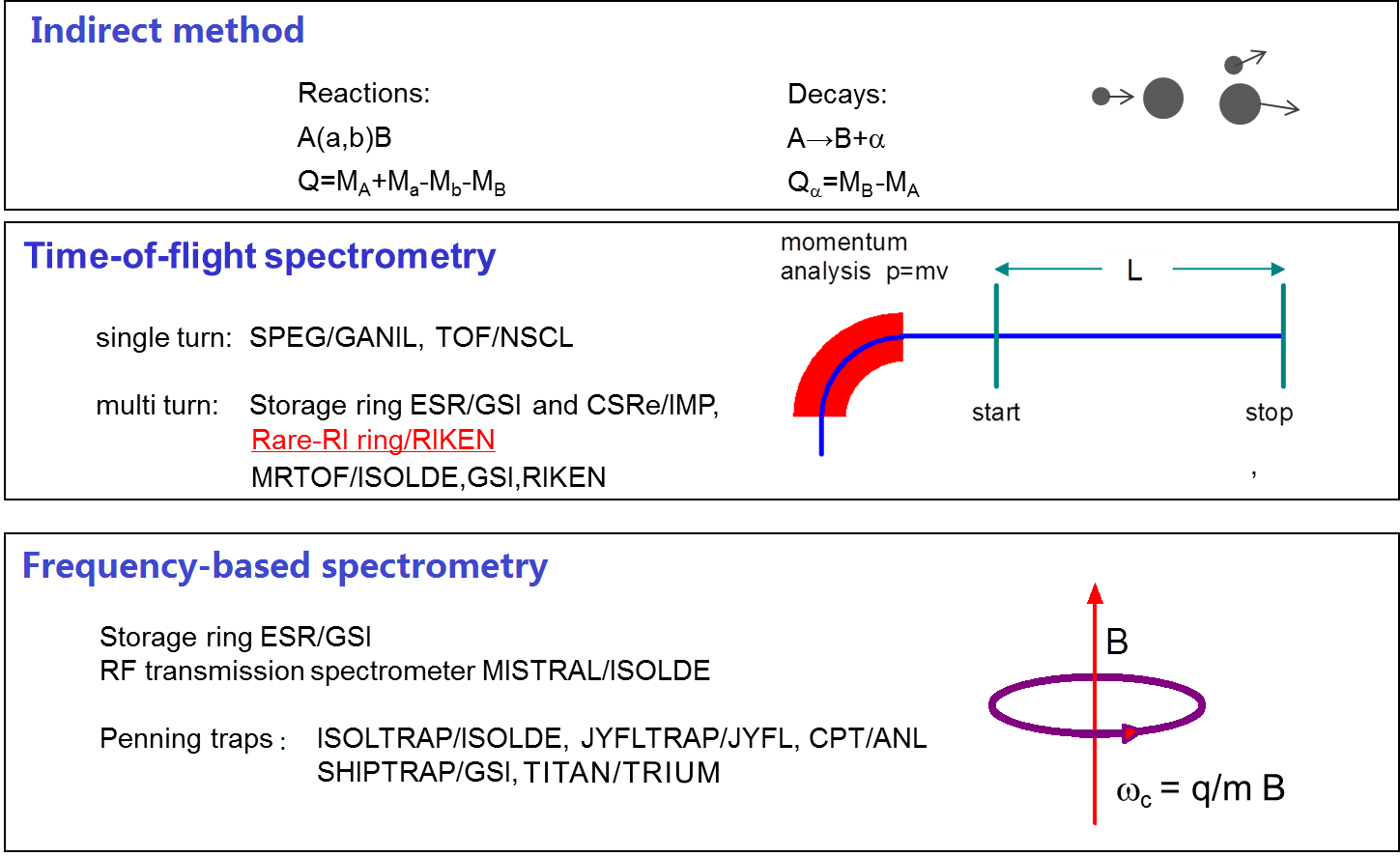}
} \caption{Schematic view of different methods for mass measurements on rare isotopes. Rare-RI ring was  commissioned successfully recently.
}
\label{fig4}
\end{figure*}

Modern experimental methods of mass measurement of rare isotopes can generally be grouped into three classes: time-of-flight mass spectrometry (TOF-MS), frequency-based spectrometry, and
indirect methods. Operational TOF facilities include the single-pass TOF spectrometers at GANIL and NSCL, and the multi-turn instruments at GSI and IMP (i.e., isochronous mass spectrometers), and the multi-reflection TOF (MR-TOF) spectrometers at GSI, CERN, and RIKEN. The frequency-based facilities are the Schottky mass spectrometer at GSI, the radio frequency (RF) transmission spectrometer MISTRAL at ISODE, and many Penning traps worldwide~\cite{Blaum2013PS152}.
The TOF and frequency-based methods are often mentioned together as direct mass measurement methods because unknown masses (in fact, mass-to-charge ratios) are directly determined by calibrators
with well-known masses. Indirect methods use nuclear reactions or decays; the unknown mass is calculated from known ones in the reaction or decays plus the determined $Q$ values. These different techniques are schematically summarized in Fig.~\ref{fig4}.

\begin{figure*}
\centering
\includegraphics[width=0.45\textwidth, angle=0]{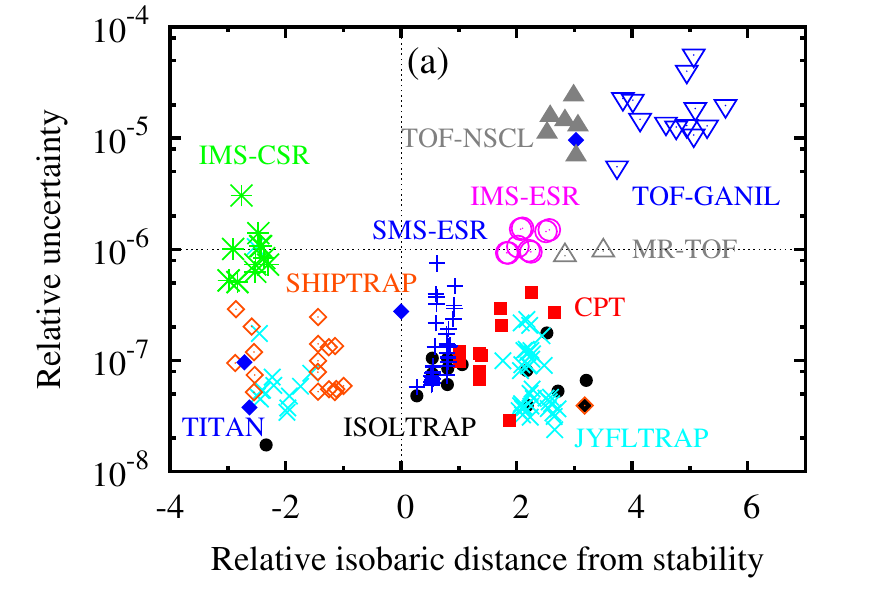}
\includegraphics[width=0.45\textwidth, angle=0]{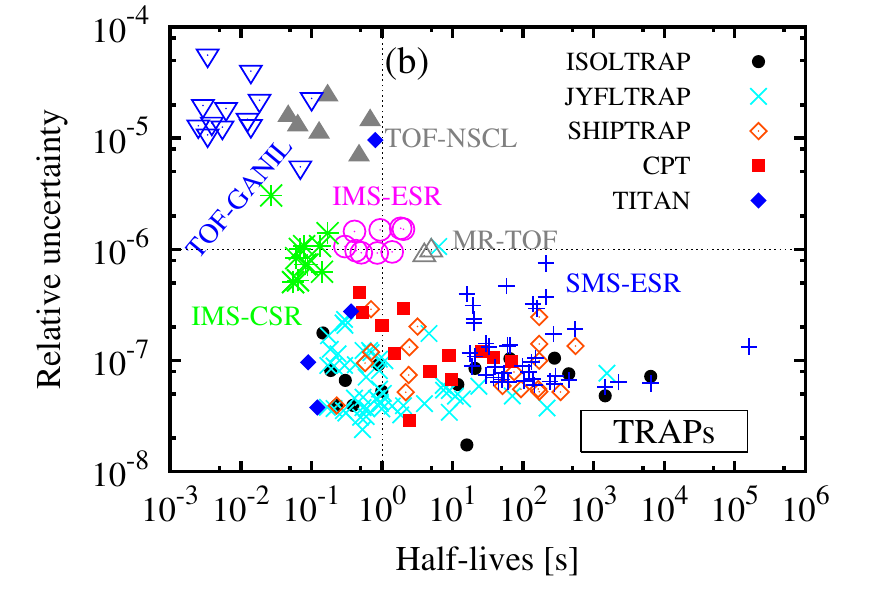}
 \caption{Precision vs. exoticity (a) and vs. half-lives (b) for different nuclear mass measurement facilities. The quantity of relative isobaric distance defined by $Z_0-Z$ is normalized by $10 A^{-2/3}$, since the dripline is reached earlier for lighter masses.
 $Z_0 = \frac{A}{1.98+0.0155 A^{2/3}}$ stands for the proton number of the most stable isotope in the isobaric chain with mass number $A$.
 Positive number and negative number in horizontal axis indicate
 neutron-rich and neutron-deficient nuclei, respectively.
 Only those nuclei with masses measured for the first time since 2003 are included in this plot.}
\label{fig5}
\end{figure*}

New measurements since 2003 are plotted in Fig.~\ref{fig5}(a) and ~\ref{fig5}(b) as  precision vs. exoticity and
 precision vs. half-lives, respectively, showing the capabilities of various direct mass measurement methods.
Among the methods, Penning traps~\cite{Blaum2006PhysRep425} and storage rings~\cite{Franzke2008MSR27} have become the flagship facilities in the journey to
weigh masses of exotic nuclei precisely over the last 10 years.
In fact, the impressive extent of the explored mass surface of nuclides, as seen in Figs.~\ref{fig-chart} and~\ref{fig5}, is owed largely to these two methods.
Penning trap mass measurement facilities~\cite{Blaum2006PhysRep425,Blaum2013PS152} exist in many nuclear physics laboratories,
whereas presently there are only two heavy-ion storage ring facilities performing mass measurements of exotic nuclei~\cite{Litvinov2013NIMB603},
experimental storage ring (ESR) at GSI (Germany) and CSRe at IMP/CAS at Lanzhou. 
In this section, we will give a short overview of various methods and their characteristics, with a focus on their advantages and limitations,
as well as possible ways to improve them toward measurements of $r$-process nuclei.

\subsection{Frequency-based mass spectrometry}

\textbf{Penning trap mass spectrometry (PTMS):} PTMS~\cite{Blaum2006PhysRep425} is the most widely used, well explored, and established technique for measuring atomic (nuclear) masses of unstable nuclei.
It is coupled with the isotope separation on-line (ISOL) method and makes measurements with the ions nearly at ``rest.'' Current PTMS uses the TOF ion cyclotron resonance (TOF-ICR) method~\cite{Blaum2006PhysRep425},
in which ionic motion is excited by applying an RF quadrupolar field at the true cyclotron frequency
$\nu_c = (1/2\pi)(q/m)B$ for ions with a mass-to-charge ratio $m/q$. $B$ is the magnetic field strength. 

PTMS is capable of precise measurements in the parts per billion range,
outperforming other methods in precision, as clearly seen in Fig.~\ref{fig5}(a). In the last decade,
Penning trap facilities have contributed not only by obtaining precise mass values for nuclides near the stability line, but, together with the storage ring facilities (to be discussed below),
providing valuable experimental results for a wealth of unstable, short-lived nuclides~\cite{Kluge2010HI}.

The achievable relative precision of PMTS with the TOF-ICR technique is
  \begin{eqnarray}
    \dfrac{\delta m}{m} &\propto & \dfrac{1}{ B \cdot T_{rf}\sqrt{N_{ion}}}\dfrac{m}{q}\;,
    \label{eq:PTMS}
  \end{eqnarray}
where $B$ is the Penning trap magnetic field strength, $T_{rf}$ is the excitation time, and $N_{ions}$ is the number of detected ions.
Typically, to achieve resolving powers of several million and a precision of about 10$^{-7}$, a trapped singly charged ion must be excited for a sufficient
amount of time before extraction, $T_{rf}$, and sufficient ions, $N_{ions} \gtrsim 100$, are needed as well to fit the resonance curve.
Considering the additional time required to produce
a rare isotope by the ISOL method and capture it within the trap, mass measurements of short-lived
nuclei become very difficult. For the ISOL method in particular, considering the delay during the process of diffusion and ionization,
the half-lives of rare isotopes that can be studied are normally limited to a few hundred milliseconds. This imposes a boundary condition on the excitation time and,
therefore, limits the achievable resolving power.
For some gases and alkaline elements, the limit may be on the order of 10 ms~\cite{TITANPhysRevLett2008PRL101-11Li}.
On the other hand,
exotic nuclei far from stability often have millisecond half-lives and minute production rates.
All these factors hamper both the achievable accuracy and speed of PTMS.
The limitation of about 0.1 s half-lives is clearly seen as the sharp boundary in the PTMS data in Fig.~\ref{fig4}(b).

Eq.~(\ref{eq:PTMS}) shows that by significantly increasing $N_{ions}$, a higher accuracy can be achieved  even for short-lived heavy nuclei.
This is possible especially for some very special cases such as $^{11}$Li ($T_{1/2}$ = 8.75 ms). To achieve 
a mass resolving power of about $\Delta m$ = 86~000 
and an excitation time of 18 ms,
the total number of $^{11}$Li ions has to approach $N=10 000$~\cite{TITANPhysRevLett2008PRL101-11Li}.
Note, however, that the number of ions $N_{ion}$ is limited by the RIB production yield as well as the efficiency of the spectrometer.
In fact, even large yields do not necessarily increase $N_{ion}$,
as PTMS measurements have to be performed with only a few ions at a time in order to exclude
or minimize systematic effects such as ion--ion interactions. Accordingly, prolonging $T_{rf}$ to increase the precision is 
practically useful only when this extension would not result in significant ion loss from radioactive decay. Neutron-rich nuclei with half-lives of a few tens of milliseconds are clearly not suitable.

Instead, several possible ways of overcoming the limitations of PTMS have been considered, e.g., by using superconducting magnets with a higher magnetic field $B$, by replacing the quadrupole excitation with multipolar excitation of ion motion, by using the Ramsey TOF-ICR technique, by using the new phase-imaging ICR (PI-ICR) technique, and by increasing the charge state of stored ions.
Although using a higher field strength has the primary advantage of shortening the measurement time directly for a given statistics and mass resolution, the magnetic field strength currently used is still limited to less than 10 T for PTMS.

The application of the multipolar technique, on the other hand, will improve the statistical uncertainty of frequency determination~\cite{PTMS2007IJMS45}
and thus the mass resolution. As a first experimental demonstration of this possibility, the octupolar technique was applied to determine the mass ratio of the
$^{164}$Er-$^{164}$Dy mass doublet~\cite{PTMS2011PhysRevLett.107.152501}. This isobaric pair differs in mass by only about 25 keV.
A resolving power of about $2\times 10^7$ was obtained, exceeding that of the conventional quadrupolar technique for the same excitation time by more than an order of magnitude.

The Ramsey TOF-ICR technique was suggested more than 20 years ago~\cite{Bollen1992NIMB490} as a possible way to improve  the precision of nuclear mass determinations. The prerequisite for the Ramsey method is an understanding of the observed TOF cyclotron resonance curves using time-separated oscillatory fields.
Its validity was recently demonstrated experimentally by measuring the mass of $^{38}$Ca [$T_{1/2}$= 440(12) ms]  using the Penning trap mass spectrometer ISOLTRAP at CERN~\cite{PTMS2007PhysRevLett.98.162501}.  
The relative uncertainty was determined to be  1.1$\times 10^{-8}$. 
 
A different approach to measuring the cyclotron frequency in a Penning trap using
the PI-ICR technique was proposed~\cite{PTMS2013PhysRevLett.110.082501}.
In this method, the cyclotron frequency $\nu_c$  in a magnetic field  can be determined 
by projecting the ion motion in the trap onto a high-resolution position-sensitive microchannel plate (MCP) detector.  Compared to the presently used TOF-ICR methods, the new technique offers a 40-fold increase in the resolving power and a fivefold gain in the precision of cyclotron frequency determination. The technique can be employed for mass measurements of very short-lived nuclides or for high-precision measurements of stable nuclides. For instance, in this method, low-lying isomeric states with excitation energy at the 10-keV level can be easily separated from the ground state.

Progress with highly charged ions (HCIs) offers a similar bright future in terms of both resolving power and accuracy.
The first measurement with highly charged states was recently performed at TITAN on a neutron-deficient $^{74}$Rb isotope ($T_{1/2}$ = 65 ms)~\cite{PTMS2011-PhysRevLett.107.272501}, where 
Rb isotopes were successfully charge-bred in an electron beam ion trap to $q=8-12^+$ prior to injection into the
Penning trap. This approach may extend the PTMS method to more exotic nuclei that were impossible to access with singly charged ions,
provided that the additional step of charge breeding can be realized in a time period in which excessive decay loss does not occur. A few challenges for HCIs remain; e.g.,  an increased electron beam current is needed to reduce the breeding time, and 
the charge breeding efficiency needs to be improved.

Nevertheless, applications of  novel techniques are helpful in reducing the measurement time and thus accessing more exotic nuclei.
High precision can be achieved even for lower production yields and/or shorter half-lives.
Most importantly, the significant increase in resolution makes them extremely suitable for ultrahigh-resolution
measurements such as fundamental symmetry studies and investigation of long-lived ($>$0.1 s) low-lying isomers with excitation energies down to a few tens of kiloelectron volts.

\textbf{Schottky Mass Spectrometry (SMS)}: SMS, first developed at the ESR at GSI, is another representative example of frequency-based mass spectrometry.
Its precision and resolution are almost competitive with those of the Penning trap, and it has the primary advantage of
a strong ability to map the nuclear mass surface using a single measurement setting. This is due to the application of an in-flight separation method and the large acceptance of storage rings.

The principle of storage ring mass spectrometry (SRMS), to the first order, can be simply
expressed as
\begin{equation}
  \frac{\Delta f}{f} = -\alpha_p\frac{\Delta(m/q)}{m/q}+\eta\frac{\Delta p}{p} \; ,
  \label{eq_SRMS}
\end{equation}
where $f$, $m/q$, and $p$ are the revolution frequency,
mass-to-charge ratio, and momentum of a circulating ion in the ring,
respectively.
Here $\Delta$ denotes the difference between two discrete values. $\Delta(m/q)/(m/q)$ is the relative difference between the mass-to-charge ratios of two ion species, and
$\Delta p/p$ is the difference between the mean velocities of the corresponding ion ensembles. $\eta = 1/\gamma^2 - 1/\gamma_t^2$ is the frequency
dispersion function or phase slip factor, where $\gamma$ is the Lorentz factor. $\gamma_t=1/\sqrt{\alpha_p}$ is the
transition point of the storage ring, at which the revolution frequency becomes independent of the energy for each ion species with a fixed $m/q$.

The relative spread $\delta f/f$ of the revolution frequency distribution of each ion component is determined only by the momentum spread $\delta p/p$ of the corresponding ion ensemble and
is given by
\begin{equation}
  \left|\frac{\delta f}{f}\right| = \eta\frac{\delta p}{p} \; .
  \label{eq_SRMS2}
\end{equation}
In contrast to $\Delta$ in Eq.~(\ref{eq_SRMS}), $\delta$ denotes the width
(usually the full width at half-maximum, FWHM) of the velocity, momentum, or frequency
distribution of each ion ensemble.

Exotic ions produced by nuclear reactions after injection into a storage ring still exhibit a considerable relative momentum
spread of up to $1\%$~\cite{Franzke2008MSR27}.  Therefore, the last term in Eq.~(\ref{eq_SRMS}) severely limits precise mass measurements.
If this term can be reduced to a relatively negligible value, the revolution frequency becomes a direct measure of the mass-to-charge
ratio~\cite{Franzke2008MSR27,SRMS2013PPNP84,Litvinov2013NIMB603,Bosch2013IJMS151}.

To minimize the term containing the momentum spread in Eq.~(\ref{eq_SRMS}),
an electron cooler can be used to force all the circulating
ions in the ring to have an identical velocity (with a velocity
spread down to about $\Delta v/v \sim 10^{-7}$)~\cite{Steck2004NIMA357}; 
thus, the revolution frequency is related to
the mass-to-charge ratio only in the first order.  Then the revolution
frequency is deduced from the fast Fourier transform of the
induced signals captured by a nondestructive Schottky probe.  

Two adjacent peaks in a spectrum can be resolved only if their separation is larger than
their full linewidth (FWHM), i.e., $ \Delta f \geqslant  \delta f$.
This is equivalent to
\begin{equation}
  \left|\frac{\Delta f}{f}\right| = \left|-\alpha_p\frac{\Delta(m/q)}{m/q}\right|  \geqslant \frac{\delta f}{f} = \eta\frac{\delta p}{p} \; .
  \label{eq_SRMS3}
\end{equation}
In SRMS, we can define the resolving power $R_q$ using $|(m/q)/(\Delta(m/q))$, as follows:
\begin{eqnarray}
  R_q &=& \left|\frac{\alpha_p}{\eta}\frac{1}{\delta p/p}\right| = \left|\frac{1}{\gamma_t^2}\frac{1}{\eta}\frac{1}{\delta p/p}\right|  \nonumber\\
   &=& \left|\left( 1- \frac{\gamma_t^2}{\gamma^2}\right)\frac{\delta p}{p}\right|^{-1} \; .
  \label{eq_SRMS4}
\end{eqnarray}

Clearly, $R_q$ is inversely proportional to the momentum spread $\delta p/p$ of the ion beams, whereas
$\delta p/p$ depends on the number of stored ions~\cite{Steck1996PhysRevLett.77.3803,Steck2003JPB36}. For low
beam intensities of about 1000 stored ions, a $\delta p/p$ value of about $5\times 10^{-7}$ has been observed. However, a higher intensity of stored ions in the ring would
suddenly increase the momentum spread by as much as one order of magnitude. Another possible way to increase the resolving power is to use secondary fragments at higher energy,
i.e., at higher $\gamma$. However, this technique is limited by the maximum cooler voltage that can be used for the electron cooler.

A lower $\gamma_t$ close to $\gamma$ is favorable for obtaining a higher resolving power. However,
the momentum acceptance of a storage ring decreases strongly with increasing $\alpha_p$. Therefore, in reality
one has to find a compromise between the ring acceptance and the desirable resolution. By increasing $\alpha_p$ to about 0.36, which is a factor of 2 larger than that in the usual mode
of the ESR ($\alpha_p$ = 0.18), the resolving power can be increased by a factor of five compared to that currently achieved at GSI.
In particular, $R_q$ goes even to infinity if the ring is operated in a special case with $\gamma_t=\gamma$, i.e., the isochronous mode. 
This is the second way to operate a storage ring as a mass spectrometer.
This model will be discussed later in this section.
Nevertheless, this may open a new method with ultrahigh precision that can be comparable to PTMS.
Considering the large acceptance of the storage ring (although it is lower than that of the standard mode), this new method will be ideal for cases such as systematic searches for unknown low-lying isomers through the nuclear chart.

The highest resolving power achieved in reality is worsened by the temporal variation of the revolution frequency, e.g.,
by instabilities in the power supplies for the SRMS magnets or the radial velocity profile of the electron beams in the
cooler. This may be addressed in two ways: the use of improved averaging
techniques during data analysis and further stabilization of the power supplies
of the ring magnets and electron cooler. The former has been realized by using the
correlation matrix method to correct the slow drift in the revolution frequency over
time~\cite{Litvinov2005NPA756,Sun2008Nucl.Phys.A812} and by taking into account the velocity profile of the cooler electrons and the residual ion--optical dispersion
in this part of the storage ring~\cite{Chen2012NPA882}.  

A recent SMS experiment mapped a new territory of 
about 150 heavy neutron-rich isotopes on the mass surface~\cite{Chen2012NPA882}. Accurate new mass values with an average precision of about 19 keV for 33 neutron-rich,
stored exotic nuclei in the element range from platinum to uranium were obtained for the first time. Among them,
five new isotopes, $^{236}$Ac, $^{224}$At, $^{221}$Po, $^{222}$Po, and
$^{213}$Tl, were discovered~\cite{Chen2010PLB234}.
The measured mass resolving power was about $1.7\times 10^6$.
Hence, isobars and even isomers with excitation energies
of down to about a few hundreds of kiloelectron volts could be resolved.
This technique even allows one to examine the nuclear reaction mechanism responsible for production
of exotic nuclei. For instance, the important role of cold fragmentation and
nuclear charge-changing reactions is revealed in this experiment
by the observation of the most neutron-rich isotopes to date in the
region of lead and uranium~\cite{Chen2012NPA882}.

Using a nondestructive Schottky pickup probe,
one can monitor the variance in the number of particles for the same ion
species and thus determine the
lifetimes of exotic nuclei. One even can trace the fate of each ion in the ring. Operation of the storage ring for such special cases is known as 
single-ion decay spectroscopy~\cite{Litvinov2011RPP74}. 
Further, SMS is an extremely sensitive technique. Often the mass of a single stored ion can be determined
with sufficient precision for nuclear astrophysics and nuclear structure physics.
This capability is essential for studying nuclides with tiny production rates.
As an example, one $^{208}$Hg nucleus was identified as a H-like ion during a two-week experiment;
this mass is crucial to addressing the proton--neutron interaction
strength around the doubly magic $^{208}$Pb nucleus~\cite{Chen2009PhysRevLett.102.122503}.
The sensitivity to single stored ions in SMS has been used to resolve low-lying isomeric states with excitation energies higher than about 100 keV~\cite{Sun2007EPJA31,Surry2010PhysRevLett.105.172501,Surrey2012PhysRevC.86.054321,Chen2013PhysRevLett.110.122502,Surrey2015PhysRevC.91.031301}.

A typical frequency spread and $m/q$ range obtained from previous SMS experiments are $\Delta f/f \sim 0.53\%$ and  $\Delta(m/q)/(m/q) \sim 3\%$, respectively.
However, the electron cooling time is directly
proportional to the corresponding velocity difference between the cooler
electrons and the hot fragments by a power of three ($t_{cool}
\propto \Delta v^3$)~\cite{SRMS2013PPNP84}.
For large spreads of hot fragments, the completion of electron cooling would require several tens of
seconds or even a few minutes. Accordingly, SMS is presently restricted to long-lived exotic nuclei ($>10$ s).
Combining electron cooling with fast stochastic pre-cooling can shorten this time span to less than 10 s~\cite{Geisse2004NPA150}.
This scheme, however, can be applied only for a fixed energy determined by the stochastic cooling mechanics, e.g., 400$A$ MeV for the ESR.

A new type of resonant Schottky detector~\cite{Nolden2011NIMA69} was installed at both the ESR and CSRe recently. This 
will support a campaign to improve all aspects of SMS, from its resolution to its sensitivity.
In comparison with the capacitive Schottky pickup used in previous SMS experiments~\cite{Litvinov2005NPA756},
this new probe can produce Schottky spectra with a significantly enhanced signal-to-noise ratio~\cite{Kienle2013PLB638}.
Accordingly, the mass resolving power can be improved by a factor of 4 owing to
the large working frequency of the new resonator. This improvement is particularly important for detection
of ions with a small charge state, a low production rate, and very similar $m/q$ values.
Its high speed was demonstrated in a pilot experiment, in which the revolution frequencies of $^{213}$Fr ions could be measured with a time resolution of only 32 ms~\cite{Sun2011GSI},
about a factor of 1000 faster than the standard pickup.
This feature enables detection of rapid changes such as short-lived nuclear decays or additional averaging toward 
obtaining a frequency spectrum with better quality.

\subsection{Time-of-flight mass spectrometry}

{\textbf{TOF-B$\rho$-MS:}} TOF-MS is a pioneering method applied to
the study of short-lived nuclei~\cite{Lunney2003RMP75} and still plays an active role in this area~\cite{MSUTOF2013IJMS145}.
TOF-MS uses a precise measurement of the flight time $t$, within which an ion travels a known flight path length $L$
along a magnetic beam line system with a fixed magnetic rigidity B$\rho$.
The mass-to-charge ratio $m/q$ is derived from the equation of motion:
\begin{eqnarray}\label{eq:TOF1}
 \dfrac{m_0}{q} &=& \dfrac{B\rho}{\gamma L/t} = B\rho \sqrt{\left(\frac{t}{L}\right)^2 - \left(\frac{1}{c}\right)^2} \; ,
\end{eqnarray}
where $c$ is the speed of light, and $\gamma$ is the Lorentz factor.

TOF-MS combines high-resolution flight time determination
with a magnetic rigidity measurement. The TOF is measured using either fast timing scintillators or MCP-based secondary electron emission detectors, whereas
B$\rho$ is measured by detecting the position of each ion at a large dispersive focus with a position-sensitive gas or MCP detector.
An energy loss detector is often used as well for particle identification.
This complementary detector is essential for accessing heavier systems, as the TOF-B$\rho$ technique itself does not allow unambiguous identification of exotic nuclides with similar $m/q$ values.
The first TOF-B$\rho$ technique for mass measurement of exotic nuclei was applied at the SPEG spectrograph at GANIL~\cite{SPEG1989NIMA509,SPEG2001}.

The mass analyzing power can be easily deduced from Eq.~(\ref{eq:TOF1}):
\begin{eqnarray}\label{eq:TOF2}
 \dfrac{m_0}{\sigma_{m_0}} =1/\sqrt{\frac{\sigma_{(B\rho)}^2}{(B\rho)^2}  + \frac{1}{k^2}\left(\frac{\sigma_t^2}{t^2} + \frac{\sigma_L^2}{L^2}\right)} \;,
\end{eqnarray}
where $k=1-(L/(ct))^2$.
Although the TOF can be determined with very high precision, the measurements of the magnetic rigidity (via position measurement) and flight length
would severely limit the precision of the resultant mass. Rather than directly determining $m$ from the experimental $t$, B$\rho$, and $L$ values according to Eq.~(\ref{eq:TOF1}),
in reality one calibrates the $t$--$m_0/q$ relationship using nuclei with well-known masses. A similar approach has been employed in SRMS.

The merits of TOF-B$\rho$-MS are well recognized as its very broad elemental and isotopic distributions resulting from high-energy projectile fragmentation or fission when it is
combined with high-resolution, high-transmission magnetic spectrometers.
Indeed, the presently running TOF-MS at NSCL~\cite{MSUTOF2012NIMA171} is nothing but an optimized in-flight separator.
This method provides, therefore, the most efficient mapping of an entire region of the
nuclear mass surface and thus is especially suitable for weighing short-lived nuclei near the drip line.
For instance, the masses of Borromean drip-line nuclei
$^{19}$B, $^{22}$C, $^{29}$F, and $^{34}$Na have been measured only by TOF-B$\rho$-MS to date~\cite{SPEG2012PhysRevLett.109.202503}.
New results from GANIL~\cite{SPEG2007PhysLettB649,SPEG2012PhysRevLett.109.202503} and NSCL~\cite{MSUTOF2011PhysRevLett.107.172503,MSUTOF2014PhysRevLett.114.022501}
are summarized in Fig.~\ref{fig5}.

A mass resolution of (2-4)$\cdot 10^{-4}$ was obtained using a combination of TOF and magnetic rigidity measurements~\cite{SPEG2001,MSUTOF2012NIMA171}.
This corresponds to about $\pm$5 MeV of the mass excess for a single nucleus with $A = 50$.
The final uncertainties governed by the statistical law, depending on the number of detected particles,
range from 100 keV for thousands of events (nuclei relatively close to stability) to 1 MeV for tens of
events (nuclei approaching the ends of isotopic chains).
As Fig.~\ref{fig5} shows, the relative mass precision is limited to about $10^{-5}$, but the fast speed is a distinct advantage.

As Eq.~(\ref{eq:TOF2}) indicates, a high-resolution time-pickup detector and position-sensitive detector are crucial for
improving the mass resolving power. At NSCL, the time resolution obtained from 1.5 cm $\times$ 2.53 cm $\times$ 0.254$^t$ mm plastic scintillators is $\sigma \approx 80$ ps
after correction for the position dependence.
Recent development of plastic scintillators with fast decay times and high-speed phototubes provide us the opportunity to make fast timing
measurements with a resolution of less than 10 ps~\cite{TOFdetector2008NIMB266,TOFdetector2011NIMA354,TOFdetector2013NIMA40}.
Independently, we have achieved a time resolution of less than 10 ps with a 30 mm $\times$ 10 mm $\times$ 3 mm$^t$ plastic scintillator~\cite{Zhao2015_Private}.
For mass measurement, such a small detector can be used at the focus plane.

The  relative magnetic rigidity can be obtained from a pair of position measurements at a magnetic beam line system with one at the dispersive focus.
A position resolution of about 300 $\mu$s is obtained for NSCL, which is similar to that of the position-sensitive gas detector used at GANIL.
Position determination using a high-gain camera with an image intensifier and position-sensitive MCP detectors is under investigation at Beihang~\cite{Tanihata2015_Private}.
Both position readouts share a similar principle: the time information on heavy ions is first passed on to the secondary electrons or light emitted from their passage through
a thin foil or phosphor. The emitted electrons or light will be guided and then focused onto an MCP or a CCD camera for position readout.
One concern for all the detectors is that they should have as small an effect as possible on the beam itself to avoid introducing any additional systematic errors.
By using augmented timing and position detectors, it is possible to improve the mass resolution by several times. 

A more efficient way to improve the mass resolution of TOF-B$\rho$-MS is to increase the total flight path length.
A longer $L$ will directly improve not only the precision $\sigma_L/L$, but also the time precision $\sigma_t/t$.
For the GANIL and NSCL setups, $L$ is 116 and 59 m, respectively. A natural extension to such a single-pass TOF spectrometer
is to develop multi-turn facilities, e.g., cyclotron-like or synchrotron-like rings, or MR-TOF systems to extend the flight path.
The flight paths of these devices are typically two orders of magnitude (up to a few tens of kilometers) larger than conventional TOF mass spectrometers.
As discussed below, this significantly increases both the mass resolving power and mass precision. 
One drawback of multi-turn facilities is the loss of beam intensity during the injection phase.

\begin{figure*}[!t]
\centering
\includegraphics[width=0.8\textwidth, angle=0]{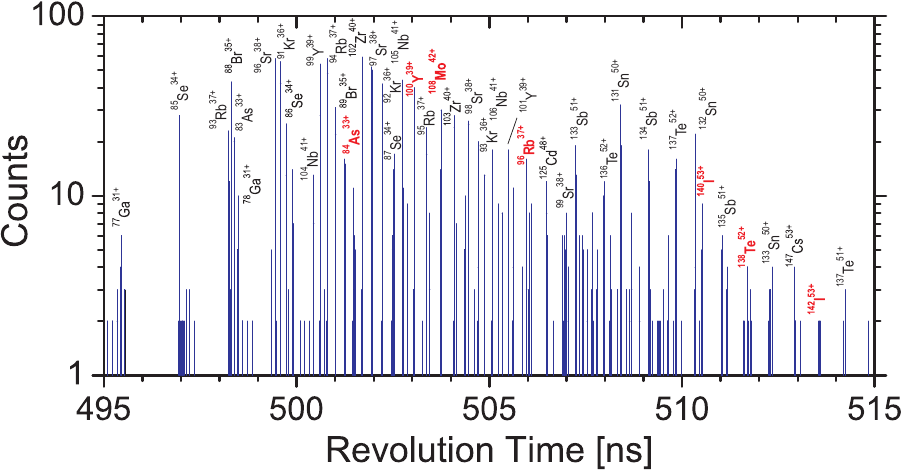}
 \caption{Revolution time spectrum measured with IMS.
Only the prominent peaks are labeled with the corresponding isotope identifications.
The nuclides with known masses (black letters) and previously unknown masses (red letters) are indicated according to the AME2003 compilation. Taken from Ref.~\cite{Sun2008Nucl.Phys.A812}.}
\label{133Sn_total}
\end{figure*}

The population of isomeric states with lifetimes on the order of or greater than the flight time through the system
(from about a few hundreds of nanoseconds to about 1 ms) is a potential problem, as the resolution of the system is not sufficient to resolve the typical mass difference between the ground and isomeric excited states.
The contribution of such states will cause a systematic shift in the measurement
toward less bound masses. One solution is to use the flexibility of TOF-B$\rho$-MS to couple it with $\gamma$-ray detectors. This was realized at SPEG at GANIL~\cite{SPEG2001,SPEG2002PhysRevC.65.044618}.
The measurement of $\gamma$ rays in coincidence with ions of interest allowed the presence of isomers to be identified and the populations to be derived.
This can provide valuable information on the isomeric energies, half-lives, $\gamma$-ray energies, and isomeric production ratios. In addition,
the isomeric contributions can be corrected further from the measured masses. This will improve the ability of TOF-B$\rho$-MS, which is limited by the resolving power.
In addition to the TOF-B$\rho$-MS in operation at NSCL, a new TOF-MS is under development at IMP~\cite{Tanihata2015_Private}. To achieve an accuracy comparable to that of storage ring spectrometry, a high-resolution detector system including fast timing detectors with a precision of about 10 ps and 
ultraprecise position detectors with a precision on the order of 10  $\mu$m is required. It has two functions: it not only can determine the masses of most exotic nuclei
directly from the TOF with B$\rho$ correction, but also can serve as an efficient separator to enhance the purification power for IMS mass measurements. It will be
the first TOF-B$\rho$-MS to run at a relativistic energy of around 400 MeV/u. This makes it possible to access even heavier mass systems than can be accessed at NSCL.

{\textbf{Isochronous mass spectrometry (IMS)}}: IMS~\cite{Hausmann2000NIMA569},
a complementary method to SMS, is a multi-turn TOF setup.
Unlike the situation for SMS, for IMS the storage ring is tuned to be an isochronous ion-optical model, 
where the phase slip factor $\eta$ vanishes in the first order [see Eq.~(\ref{eq_SRMS})].
In other words, the velocity dependence is overcome by
matching the ion-optical mode of the storage ring and
the beam energy such that the velocity difference
between two particles of the same species is
counterbalanced by the corresponding change
in the orbit length. This requires that $\alpha_p = 1/\gamma^2$, the so-called isochronicity condition.
Because no cooling is required, IMS is suitable for accessing nuclei
with lifetimes as short as a few tens of microseconds. The mass of nuclides of interest can be determined
by directly measuring the flight time of the ions in the ring with a dedicated
time-pickup detector, an MCP-based secondary electron emission detector~\cite{ESR1992NIMB455,CSRe2010NIMA109,CSRe2014NIMA755,CSRe2014NIMA756}. 
Currently, IMS has been developed for the ESR (GSI) and CSRe (IMP).

Taking the ESR as an example, for a typical $\gamma$ of 1.4 ($\beta=0.7$), the isochronicity condition requires an $\alpha_p$ of
about 0.5, nearly a factor of 3 higher than that in the standard setting of the ESR.
In pilot IMS experiments~\cite{Hausmann2000NIMA569,Stadlmann2004PLB27}, the best
mass resolving power of about 110~000 and a mass accuracy of
100--500 keV were achieved. Nuclides with half-lives down to 50 ms
were observed. Further investigations revealed that the isochronous condition is strictly fulfilled only for an $m/q$ range
in a narrow B$\rho$ range. The further the system is from the ideal isochronicity, the worse the mass resolving power becomes. Investigations on the isochronicity of storage rings
are also under way to obtain the best mass resolving power~\cite{Dolinskii2007NIMA207,Litvinov2013NIMA20,CSRe2014NIMA53}. 

As a compromise, strong restriction criteria had to be applied in the analysis of the broadband IMS spectrum: only
those revolution time peaks within the so-called isochronous window, a small part of the measured revolution time spectra, were used in the
final mass determination~\cite{IMS2004PHD}.
With the successful commissioning of the Cooler Storage Ring (CSR) at the Heavy Ion Research Facility in Lanzhou in 2008, the masses of short-lived nuclides have been accurately measured by IMS using
projectile fragmentation of $^{78}$Kr, $^{86}$Kr, $^{58}$Ni, and $^{112}$Sn beams.
A typical precision of less than 50 keV was obtained by confining the data analysis to the ``isochronous'' part~\cite{Tu2011PhysRevLett.106.112501,Tu2011NIMA213}.
Masses of 16 neutron-deficient nuclides were obtained for the first time~\cite{IMP2013IJMS-review}.
However, to take the advantage of the broad $m/q$ spectrum, one has to correct the ``non-isochronicity'' effect while retaining good resolving power and precision.

One solution is to restrict the velocity or magnetic rigidity of each fragment before its injection into the ring.
This concept was realized in an experiment where the high resolution of the fragment separator is used to determine the B$\rho$ value of the injected fragments within
$1.5\cdot 10^{-4}$ at the second dispersive focal plane via a modified slit system~\cite{Geissel2006HI173}.
This novel extension has been demonstrated in a short test run~\cite{Geissel2006HI173} and
also applied in a production run for fission fragments~\cite{Sun2008Nucl.Phys.A812}.
In measurements of short-lived uranium fission fragments, more than 120 peaks in the revolution times
were recorded, as shown in Fig.~\ref{133Sn_total}. This corresponds to a typical frequency spread $\Delta f/f$ of $\sim6\%$
and an $m/q$ range $\Delta(m/q)/(m/q)$ of $\sim12\%$.
After correction of the overall drifts in the revolution times caused by magnetic field instability during the measurement for all the different ion species, a mass resolving power of about 200~000 has been achieved over nearly the entire
revolution time spectrum. The relative uncertainties amount to 
about $10^{-6}$. The above B$\rho$ selection method, however, incurs a cost of significantly reduced transmission.
Moreover, although  more than one hundred TOF peaks have been obtained, as seen in Fig.~\ref{133Sn_total}, about half of them could not be uniquely assigned to one species owing to the limited resolution
and thus have to be excluded from the present data analysis. For example, the peak labeled $^{108}$Mo$^{42+}$ could be an unresolved mixture of $^{108}$Mo$^{42+}$ and $^{54}$Sc$^{21+}$. In reality, only 71 different peaks were unambiguously identified in the accumulated TOF spectrum and were analyzed further.

A better method than B$\rho$ tagging is to make a fast event-by-event velocity correction for each ion using a high-resolution magnetic separator. As seen in Eq.~(\ref{eq_SRMS4}),
this will result in an improved resolving power for IMS without limiting the acceptance of the ring, and
the achievable resolution depends on the precision of the velocity determination.
This concept is under commissioning at the Rare-RI RING (R3) storage ring at RIKEN~\cite{Ozawa2012PTEP,RareRI2013IJMS240}.
Three TOF detectors, two at the entrance of the ring and one in the ring,
will be used to determine the velocity of each ion. The goal is to determine the velocity with a precision better than $10^{-4}$.

An alternative method of velocity correction is to develop in-ring velocity measurement for each ion~\cite{Geissel2006HI173}.
This can be realized by employing two time-pickup detectors installed in a straight section of the ring.
The first commissioning experiment was done this year at CSRe (IMP)~\cite{IMP2013IJMS-review}.
A velocity precision of about $10^{-3}$ might be possible. One drawback of in-ring velocity measurements is that
the total number of turns of circulating ions will be reduced by the energy losses in the two time-pickup detectors.

Similar to the case for SMS, turn-by-turn detection of circulating ions in IMS opens the possibility of obtaining information on the lifetime.
A pilot work in this direction was done on a 17 $\mu$s core-excited isomer in $^{133}$Sb~\cite{Sun2010PLB688}.  The excitation
energy of 4.56(10) MeV and survival time of the isomer in the ring
were determined on the basis of precise revolution time measurements
of a few individually stored fully ionized ions. The extended
in-flight half-life of the bare ions in the ESR is due to
exclusion of strong internal conversion~\cite{Sun2010NPA476c}.

A digital scope with a very high sampling rate (about 40 GS/s) is currently employed to record the timestamps of each passage of the time-pickup detector. However, the typical time needed solely to save the raw data and transfer it to disk is many times longer than the measurement time itself.
Improvement toward fast data recording and transfer
will be very helpful for using the precious beam time more efficiently. Reducing 
the beam cycle time of the main synchrotron will have a similar effect.

As mentioned before, the revolution frequency is detected using a dedicated TOF detector. Considering the
revolution frequency of about $2\times 10^6$ Hz at both the ESR and CSRe, the number of ions stored in the ring has to be limited to several tens.
Otherwise, the gain of the detector will decrease quickly because of the saturation effect, and this will significantly reduce the detection efficiency~\cite{CSRe2014NIMA756,CSRe2014NIMA755}.
In the IMS technique, at present most of the stored ions are lost in the ring during the first 1 ms of the measurement. This can be understood
as arising mainly from the nonradioactive losses~\cite{Sun2010PLB688}, in particular the
energy loss and charge-exchange processes during multiple passages
through the detector foil at each turn and the interaction
with residual gas atoms in the ring.

\begin{figure*} 
\centering
\includegraphics*[width=0.8\textwidth]{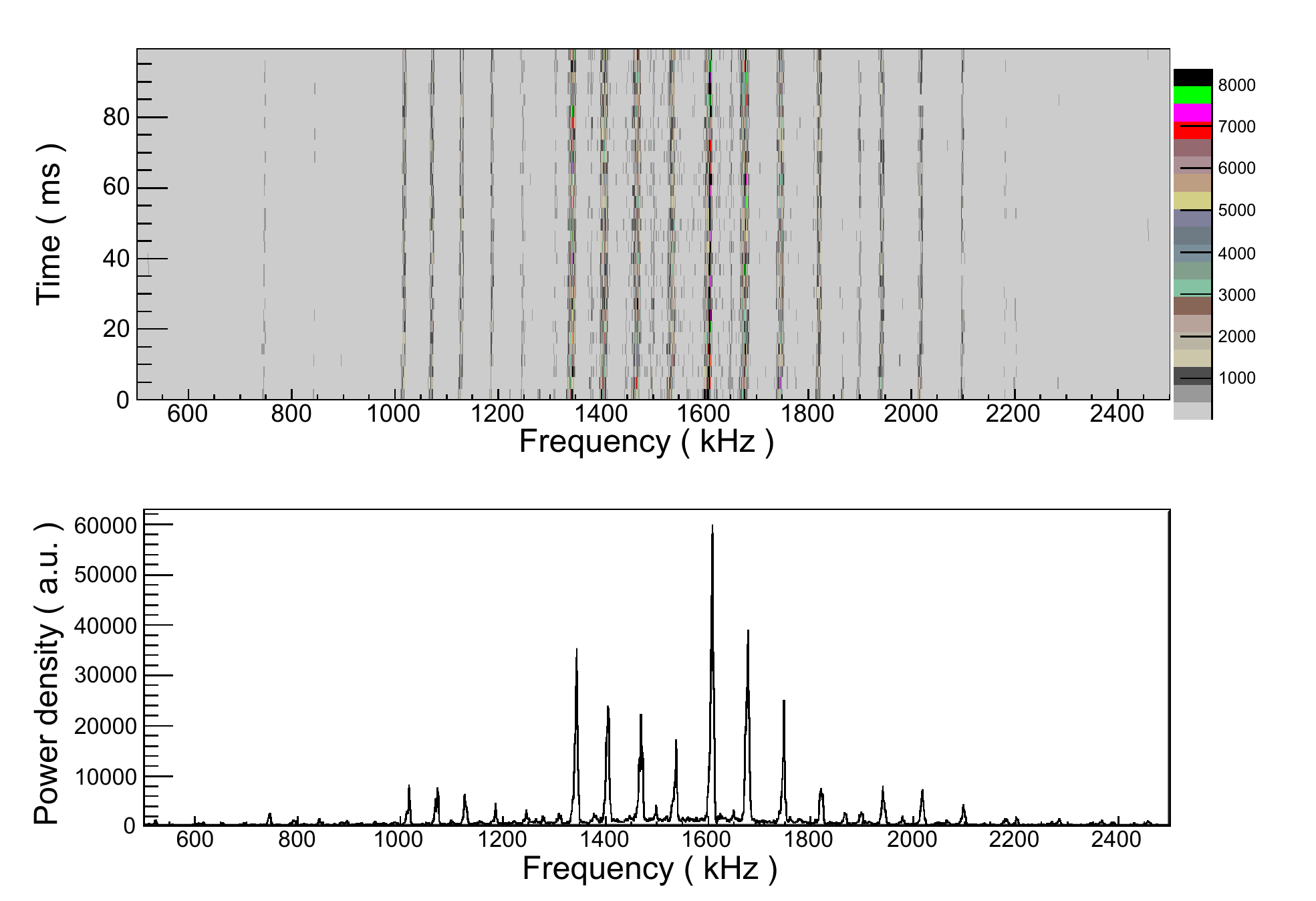}
\caption{Time evolution of the Schottky revolution frequency
spectrum (upper panel) and the projection of the first 16 ms measurement (lower panel). Here,
the raw spectrum is recorded by the new resonant Schottky pick-up. Taken from Ref.~~\cite{Sun2011PoS}. }
\label{fig:RSA}
\end{figure*}

In this respect, a nondestructive detector for IMS would have a clear advantage. This has become possible because of the fast, sensitive resonance Schottky pickup mentioned above~\cite{Sun2011PoS}.
Fig.~\ref{fig:RSA} shows a measured revolution frequency
spectrum of $^{238}$U projectile fragments centered at $^{213}$Fr,
where the ESR is operated in the isochronous mode with a
transition point $\gamma_t$ of 1.4.

The potential of applying such a probe appears clearly in the Schottky spectrum. For the first
time, it is possible to trace the fate of each circulating ion in the
ring to a time accuracy of milliseconds, e.g., 3.2 ms in the upper panel of
Fig.~\ref{fig:RSA}, which is more than one order of magnitude smaller than
the accuracy  of the standard detector. Especially for
isochronous mass measurement, it is even possible to reveal the
dispersion of the momentum, $\delta p/p$, of ions of interest by that of the
corresponding revolution frequency, $\delta f/f$, according to $\delta
p/p = 1/\eta$ $\delta f/f$.
For the 16 ms spectrum shown in the lower panel of
Fig.~\ref{fig:RSA}, a frequency resolving power of 50~000 is
obtained without any further corrections.

It is therefore very promising and interesting to further investigate
 the application of resonance Schottky detectors in IMS. One concern is that
different Schottky bands will probably overlap with each other owing to the high
working harmonic and broadband spectrum of IMS. In such cases, unambiguous identification
of the frequency spectra will become complicated. One solution could be to employ two such detectors operating at different bands. 

Numerous storage rings have been planned worldwide~\cite{Litvinov2013NIMB603,SRMS2013PPNP84}. Among them, the R3, dedicated to mass measurement of $r$-process nuclides,
will take advantage of the highest $^{238}$U beam intensity in the world (about 10 pnA at present). It is designed to have simultaneous online particle identification, precision beam emittance measurement with the fragment separator, and precision mass determination in the cyclotron-like ring. 
Last month, the R3 team performed a machine study of the
R3 using a $^{78}$Kr beam and succeeded in injecting a single particle
into the R3 and extracting it from the R3 within 1
ms. 
In China, the High Intensity Heavy Ion Accelerator Facility (HIAF), the main component of which is a multifunction storage ring system, is planned for heavy-ion-related research~\cite{Yang2013NIMB263}.
One of the storage rings is designed for short-lived nuclear mass measurements. In principle, SRTM can also be used as an extremely
sensitive isobaric separator. The plans for HIAF include this possibility for beam quality improvement through stochastic
pre-cooling and electron cooling before the beams are extracted and delivered to the secondary reaction target.

\textbf{MR-TOF-MS:} Another way to extend the flight path is to use MR-TOF-MS.
This has been realized by using electrostatic ion mirrors instead of large-scale magnets. As a result,
flight paths of several kilometers in an MR-TOF mass spectrometer can be folded into table-top dimensions.
The possibility of using MR-TOF-MS for direct mass measurements of very short-lived, exotic nuclei was noticed more than a decade ago~\cite{MRTOF2001HI531}.
It was only recently, however, that the first mass measurements of short-lived nuclides were successfully performed~\cite{MRTOF-ISOTRAP2013Nature,MRTOF-RIKEN2013PhysRevC.88.011306}.

The mass resolving power of an electrostatic multiple-pass TOF~\cite{MRTOFGSI2015NIMA172} mass spectrometer is given by
 \begin{eqnarray}\label{eq:MRTOF}
 \dfrac{m_0}{\sigma_{m}} =  \dfrac{t_0/N_a+t_a}{2\sqrt{\left(\frac{\Delta t_0}{N_a}\right)^2+(\Delta t_a)^2}} \xrightarrow{N_a\rightarrow\infty} \frac{t_a}{2\Delta t_a} \;.
\end{eqnarray}
Here, $t_0$ is the TOF from the start position to the detector without any flight path extension, i.e., the time for a single pass of the system, and $t_a$ is the overall TOF. $N_a$ is the number of passes. $\Delta t_0$ and $\Delta t_a$ are the
TOF uncertainties due to the initial conditions and aberrations in each turn, respectively.
Clearly, the overall mass resolving power tends toward $t_a/(2\Delta t_a)$ with increasing number of passes $N_a$. Eventually, the mass resolution can be
orders of magnitude higher than the resolving power achieved in single-pass TOF-MS.

MR-TOF-MS combines the advantages of conventional TOF-B$\rho$-MS---a short measurement time (milliseconds), large mass range, and very high sensitivity---with high resolution and separation ability.
It can used as an efficient and sensitive isobar separator to enhance the purification power with a resolution in excess of $10^5$, about one order of magnitude higher than that of the TOF-B$\rho$ technique.
This device can likewise be used for precision mass measurements as a competitive alternative to IMS for short-lived nuclei and to PTMS for high resolutions.
A measurement time down to several milliseconds, which includes the slowing-down time of relativistic ions, the extraction time from the gas cell, and the injections phases, allows MR-TOF-MS to access short-lived nuclei
with half-lives of the same order. Moreover, its small scale also adds to its flexibility, making it sometimes the only solution owing to limited space in the laboratory.

In the last few years, these compact devices have been commissioned at GSI~\cite{MRTOF-GSI2008NIMB4560,MRTOF-2013GSI-NIMA134}, RIKEN~\cite{MRTOF-RIKEN2014NIMA39}, and ISOLDE/CERN~\cite{MRTOF-ISOTRAP2012NIMA82}.
One of the first online mass measurements was performed at RIKEN using $^8$Li$^+$ ($T_{1/2}$ = 838 ms). A mass resolving power of about 167~000 was achieved within 8 ms for $^8$Li$^+$, which is equivalent to
that of a Penning trap with a magnetic field strength of 11 T. MR-TOF-MS was first applied to RIBs of unknown mass at CERN~\cite{MRTOF-ISOTRAP2013Nature}.
The  masses of exotic calcium isotopes $^{53}$Ca and $^{54}$Ca were determined for the first time.
At GSI, MR-TOF-MS was used for the first time for isomer-resolved studies~\cite{MRTOF2015PLB137}.
An excitation energy of $1472 \pm 120$ keV and isomeric-to-ground-state ratio of
$2.5 \pm 0.8$ were determined from the measured mass spectrum.
The achieved mass resolving power is about 250~000 within a TOF of 8.7 ms.
These pioneering experiments open up new perspectives for cases such as superheavy element research.

Like PTMS, MR-TOF-MS enables precision experiments with ions almost at rest. Accordingly, to take advantage of the in-flight method of fast and universal production, ions
of interest will have to be slowed down first and then thermalized in a gas-filled cryogenic stopping cell before they are injected into the MR-TOF mass spectrometer. 
However, this also makes MR-TOF-MS easy to couple with other techniques, e.g., for decay studies after implantation of ions of interest~\cite{MRTOF2015PLB137}.

\subsection{Indirect methods}

Indirect mass measurement via nuclear reactions or decays has long served as an important supplement to the direct methods.
Decay energy measurements~\cite{Roeckl2013IJMS349-Qvalue} represent the ``mass links'' between the atomic mass of the initial and final nuclei involved in the transition.
This enables one to deduce the atomic mass of the two states, provided that one of them and the decay $Q$ value are known.
Currently, the $\alpha$-decay and proton-decay $Q$ values still play a major role in determining the masses of superheavy elements and proton emitters, respectively.
The very intriguing case of $^{94m}$Ag demonstrates the interplay between mass determinations and decay studies. $^{94m}$Ag (21$^+$) was assigned to be the only case
where a state decays by one-proton~\cite{94mAgPhysRevLett.95.022501} and two-proton emission~\cite{94mAg2006Nature}. However, subsequent precision mass measurement by PTMS
suggests that the one-proton and two-proton decay may stem from two different isomeric states~\cite{94mAgPhysRevLett.101.142503}. Using the precise proton-decay energy of $^{53}$Co$^m$,
a recent work redetermined the mass of $^{52}$Fe~\cite{CIAE2015PRC91}.

Unlike these two types of decay, which involve only discrete transitions,
$\beta$ decay is more complicated because of its continuum energy spectrum. The $\beta$ energy has to be determined using the so-called $\beta$ endpoint measurements.
In this method, one measures the $\beta$-decay energy spectrum and then extrapolates
the $\beta$ spectrum to the $\beta$ endpoint, where the neutrino is at zero energy.
However, because of the possibly unknown level scheme and decay branches of the relevant nuclei,
the determination of the endpoint energy in most previous studies was accompanied by a large systematic uncertainty.
It is almost impossible to estimate the possible errors embedded in this method except by using other independent measurements by different techniques.
As already shown in Fig.~\ref{fig-chart}, an increasing number of new experiments reveal a shifted mass surface from $\beta$-endpoint measurements of the most neutron-rich nuclei.
Although for decades the measurement of $\beta$-endpoint energies was the only method available to provide the masses of most exotic nuclei, 
its role is becoming less important owing to the rapid development of many direct mass measurement facilities worldwide.

Reactions that exchange energy or nucleons can be used to measure the energies of binding and excitation via 
$Q$-value measurements or missing mass measurements, whereas the invariant mass method can be used to study unbound states. 
Two-body transfer reactions are generally used because it is relatively easy to reconstruct the kinematics.
Moreover, the large cross sections of transfer reactions make them also suitable for measurements at low beam intensities.
In particular, owing to recent developments in new precision detection techniques and also target techniques,
high-resolution measurements of reactions such as (p,d), (p,t) can now be made using hydrogen targets. 
Exotic nuclei can be produced by bombarding a hydrogen target with a neutron-deficient RIB.
The only prerequisite is high-resolution detectors to fully reconstruct the kinematics of the relevant reactions.
One pioneering experiment weighed the halo nuclide $^{11}$Li via the $^{11}$Li(p,t)$^9$Li reaction~\cite{Roger2009PRC79}.
This was achieved using an active target technique with a two-dimensional charge-projection and
one-dimensional time-projection chamber.
A precision as low as 100 keV can be obtained in case of good statistics of a few thousand ions.
Likewise,  the (d,p) transfer reaction or the inverse reactions
can be employed for more n-rich nuclei. 
As an example, the $Q$ value of 1.47 $\pm$ 0.02 stat. $\pm$ 0.07 sys. MeV was determined using the d($^{82}$Ge,p)$^{83}$Ge reaction~\cite{Thomas2005PRC71}. It is
in very good agreement with the updated values of 1.408 $\pm$ 0.003 MeV from a recent Penning trap mass measurement.
This method can be extended to measurements of masses of very short-lived nuclei using low-intensity beams  (down to 100--1000 pps).

For nuclei beyond drip lines, the reaction method is probably the only way to measure the associated resonance energies~\cite{Penion2001HI132,Simon2013IJMS172}
owing to the prompt destruction of the unbound system. When this method is combined with active target techniques, 
complete reconstruction of the reaction kinematics is now possible. For example, the existence of the $^7$H nuclear system was investigated via a one-proton transfer reaction with a $^8$He
beam at 15.4 A MeV and a $^{12}$C gas target~\cite{7HPhysRevLett.99.062502}.

\section{A few remarks on various methods}\label{sec:remark}

\begin{figure}[htb]
\centering
\includegraphics*[width=0.45\textwidth, angle=0]{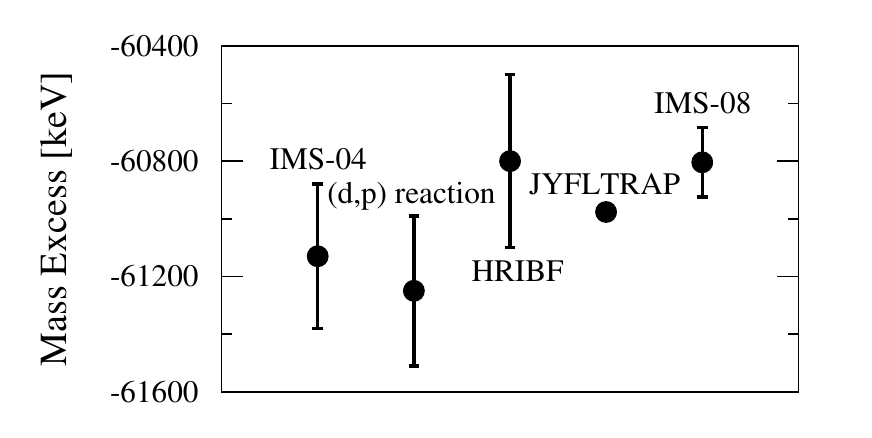}
\caption{Comparison of mass values of $^{83}$Ge determined by four different methods.}
\label{fig-83Ge}
\end{figure}

We start this section by comparing a special case in which the mass of the same nuclide is measured by different techniques.
Fig.~\ref{fig-83Ge} shows the mass data of $^{83}$Ge determined by four different methods: IMS measurement (IMS-04~\cite{IMS2004PHD} and IMS-08~\cite{Sun2008Nucl.Phys.A812}),
the Penning trap JYFLTRAP~\cite{JYFLTRAP2008PhysRevLett.101.052502}, a fast new technique developed at the Holifield Radioactive Ion Beam Facility (HRIBF)~\cite{HRIBF2006IJMS119_83Ge}, and the $Q$-value method~\cite{Thomas2005PRC71}.
In the method at the HRIBF, mass differences can be measured as position differences between known- and unknown-mass isobars dispersed at the image
of the energy-analyzing magnet. All the measurements agree with each other within one standard deviation, whereas the JYFLTRAP uncertainty is about one order of magnitude smaller.

It is always valuable and essential to have different techniques for independent checks, as the systematic error in each technique is difficult to isolate by itself.
Another reason to have such cross-checks is the need for the most reliable and accurate reference masses for direct mass measurements.
In fact, for the most n-rich nuclides, it is always a difficult  task to find reliable calibration masses, as the 
majority of the known masses of the most n-rich nuclei with $Z<28$ still have uncertainties of several hundred kiloelectron volts. The lack of references is especially crucial for
broadband methods such as IMS and TOF-B$\rho$-MS. The impact of references was demonstrated in a previous IMS experiment. With more reliable references included in the data analysis,
the mass uncertainties can be reduced by about 20\%~\cite{Sun2008Nucl.Phys.A812,IMS2008PHD}.

Another difficulty in mass measurements of n-rich nuclides is the possible coexistence of the ground state and unresolved low-lying isomers,
especially when the nuclei of interest are produced by projectile fission and separated further by the in-flight technique. The coexistence of both the ground state and an isomer state is similar to the reality in $r$-process scenarios, where low-lying isomers ($<$100 keV) can be thermally populated along with the ground states. Moreover, a new branching may open in the matter flow
when such an isomer exists along the path. In such cases, it is necessary to know the excitation energies or masses of such isomers with good precision. This would result in
high requirements for the selectivity and resolution of mass measurement techniques.

Overall, several issues have to be addressed for precision mass measurements of n-rich nuclei: the production rate, mass resolution, sensitivity, mapping power, and systematic errors.
All of these issues, except for the production rate, are clearly not fully independent. For example,
 the single-ion sensitivity makes mass determination possible even when the nuclei of interest cannot be
produced in sufficient amounts; however, it will not help to pin down the mass value in cases of poor resolution.
Given so many different techniques, each with favored regions of application, we attempt here to sort these techniques in terms of the above serious challenges. Only the direct methods are considered here.

\begin{itemize} 
  \item \textbf{High energy vs. low energy}:
  The representative methods running at high energies (typically more than a few hundreds of mega-electron volts per nucleon) are SRMS (SMS and IMS) and TOF-B$\rho$-MS,
  whereas those at low energies (almost at rest) are PTMS and MR-TOF-MS. These relevant energies define the advantages and limitations of each technique. For instance, coupling with in-flight separation gives
  SRMS and TOF-B$\rho$ powerful mapping abilities for exotic nuclei, whereas the natural extension of the ISOL method to PTMS and MR-TOF-MS provides a more purified secondary beam and high resolving power.
  
  At relativistic energies, medium-mass and heavy exotic nuclei are present to be highly charged or even fully ionized.  
  It has been verified that the observed half-lives of highly charged nuclei can differ from those in the neutral cases~\cite{Litvinov2011RPP74}, and nuclear isomers can survive much longer if the electron conversion branch is suppressed~\cite{Litvinov2003PLB573,Sun2010PLB688}. The presence of highly charged states may also have a crucial effect in $r$-process investigations, as stellar nucleosynthesis
proceeds at high temperatures, so the atoms involved are highly ionized.
 On the other hand, the use of HCIs can also significantly improve the resolving power of PTMS, and it is therefore favorable as long as the loss in efficiency caused by charge breeding is small.
 In contrast to high-energy techniques, operation of MR-TOF-MS and PTMS at low energy makes them ideal for fusion--evaporation reactions and thus for weighing trans-uranium and superheavy elements.
 
To exploit the superiority of the in-flight separation methods and PTMS or MR-TOF-MS, the ``hot'' nuclei of interest produced by projectile fragmentation or fission must be cooled down first, e.g., using a gas cell, to produce a low-energy ion beam. This has been realized at, e.g., GSI and RIKEN.

 \item \textbf{Resolving power}: PTMS has achieved the best resolving power and mass precision, and this leading position can be further ensured by the application of novel techniques such as highly charged breeding and multipolar excitation techniques. Further, breakthroughs for other techniques are now possible.
They include SMS operated in isochronous mode and MR-TOF-MS with more passages. There are in principle no special difficulties that would prevent either of these methods from realizing a resolving power
as high as that of PTMS. 

Considering the broadband characteristics in particular, SMS and MR-TOF-MS can provide a very efficient way to improve the precision of the current mass surface and contribute to cases such as searches
for long-lived low-lying isomers, which cause problems in mass determination of n-rich nuclides, and fundamental studies. Previously such studies requiring ultrahigh resolution were possible only with PTMS.

Currently, a mass resolving power of about 200~000 has been obtained for IMS, which can be further improved considerably by measuring the velocity of each stored ion for non-isochronicity correction.
With augmented timing and position precision, the resolving power of TOF-B$\rho$-MS may be increased as well.

 \item \textbf{Lifetime restriction:} With lifetimes shorter than 100 ms, which is typical for most n-rich nuclei, TOF-B$\rho$-MS, IMS, and MR-TOF-MS will be the main players for the $r$-process nuclides. Among them, TOF-B$\rho$-MS and IMS are particularly fast, and their capabilities are limited only by the flight time and the need to complete particle identification.
All current and future TOF-B$\rho$-MS and IMS instruments are, or will be, located at in-flight facilities.  A clear advantage of in-flight separation is the short flight time between the production target and the exit of the separator. This time typically ranges from a few hundreds of nanoseconds to a microsecond, which is ideally suited for studying the most exotic nuclei with the shortest lifetimes.  
 SMS, however, can be ruled out for investigations of exotic nuclei with lifetimes shorter than 1 s because of the long electron cooling time it requires.  
 
 \item \textbf{Mapping power:} Although the mapping power has been mentioned before, we feel that it is worth making a special note.
The best performance in this regard is exhibited by TOF-B$\rho$-MS, IMS, and possibly also MR-TOF-MS operated in the broadband mode. Although the precision may not be comparable with that of PTMS, these techniques
are most likely to provide the first valuable mass data on $r$-process nuclei. In comparison, Penning traps can be optimized and tuned in an experimental cycle on a
single isotope only.

 \item \textbf{Time resolution:} A unique characteristic of SMS is that one can trace the fate of each species in the ring by recording the Schottky spectrum. This allows one to explore the lifetimes of
exotic nuclei in addition to determining the nuclear masses. Moreover, nuclear decays in the ring, which are represented by frequency shifts, can provide a cross-check of particle identifications~\cite{Litvinov2005NPA756,Sun2009CPC33}. Equipped with a new Schottky probe, it is now possible to trace circulating ions in the IMS.

\end{itemize}

In addition, existing and next-generation facilities face challenges because of the overwhelming amount of contaminants produced together with the n-rich nuclides of interest.
The increasing intensity of energetic RIBs will make it possible to measure extremely exotic nuclei.
Improved purity of the nuclei of interest will not only help to eliminate many of the reaction products
that are not of interest and to reduce the total counting rate to acceptable levels in the detector, but also enable more efficient measurement and thus shorten the total beam time request.

Whereas the ISOL technique provides excellent beam intensities for a number of elements and superior beam qualities in terms of the phase-space density, the in-flight technique allows fast, chemistry-independent separation.  Unfortunately, the method of production via nuclear reactions results in a separated beam with a large phase space. This requires spectrometers with a large angular acceptance, which limits the intrinsic accuracy with which the mass might be determined. As demonstrated by IMS, this can be partially compensated by coupling the in-flight facility to storage rings.
A straightforward method of further beam purification at in-flight facilities
is to apply either a B$\rho$-TOF-$\Delta$E or two-stage magnetic separation in combination with a degrader system (the B$\rho$-$\Delta$E-B$\rho$ method).

Taking storage ring experiments as an example, in addition to the revolution time or TOF measurement, the velocity and energy loss measurement
will provide clear charge identification, as it scales with $Z^2$. When this $Z$ information is available, there should be no further confusion about the $^{108}$Mo$^{42+}$
and $^{54}$Sc$^{21+}$ (unresolved by IMS~\cite{Sun2008Nucl.Phys.A812}), as their energy losses differ by a factor of four.
We note that a determination of $Z$ in a ring was recently tested and applied further to resolving $^{51}$Co$^{^{27+}}$ and  $^{34}$Ar$^{^{18+}}$~\cite{CSRe2014PLB327}, which have
a relative $m/q$ difference of about $5\cdot 10^{-6}$. In this particular example, the carbon foil in the time-pickup detector works as an effective $\Delta E$ detector. 
Its total thickness amounts to roughly 300 turns $\times$ 19 $\mu$g/cm$^2$/turn = 0.57 mg/cm$^2$;
thus, the peaks of $^{51}$Co$^{27+}$ and $^{34}$Ar$^{18+}$ ions could not be completely separated from each other in the signal height distribution.
A dedicated, sufficiently thick $\Delta E$ detector will certainly be helpful.
Other methods have been proposed, e.g., using an individual injection system based on fast online particle identification with a high-resolution and large-acceptance magnetic fragment separator~\cite{Ozawa2012PTEP}, or using Cherenkov light detection as a velocity selector~\cite{Yamaguchi2014NIMA123}. 

\begin{figure*}
\centerline{
\includegraphics[angle=0, width=0.40\textwidth]{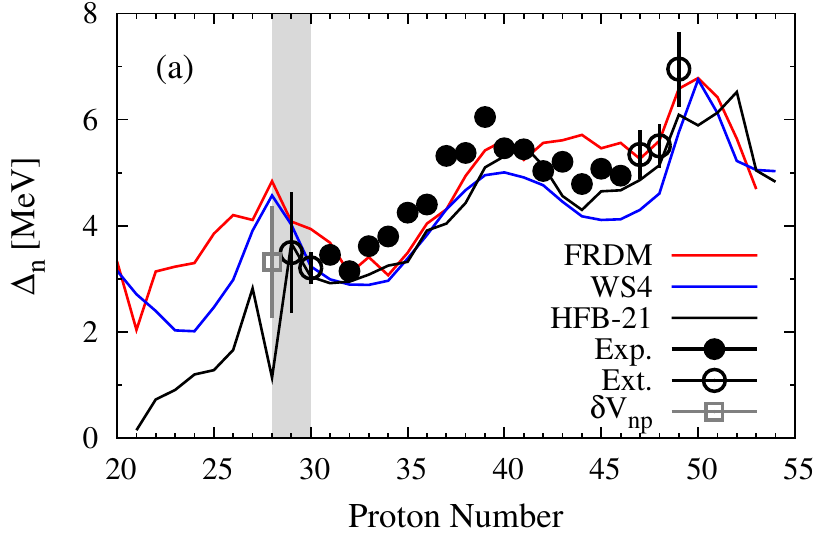}
\includegraphics[angle=0, width=0.40\textwidth]{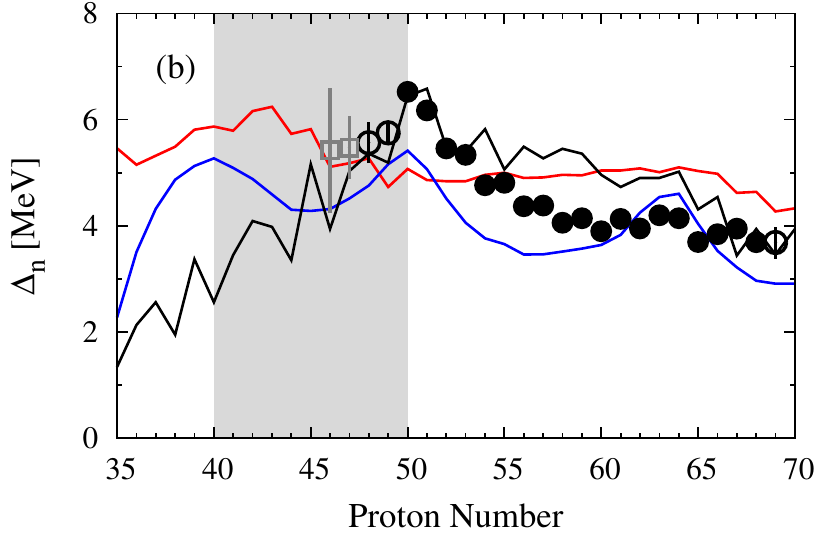}
} \caption{Experimental neutron shell gaps (full dots) for $N = 50$ (a) and 82 (b).  
 The predictions in the FRDM, WS4, HFB-21 and $\delta V_{np}$ models (open squares) and the extrapolated values (open circles) in the AME2012 are also shown for comparison.
  The nuclei in the shadowed areas are expected along the $r$-process paths. }
\label{fig:nshell}
\end{figure*}

\section{Conclusion} \label{sec:conclusion}

Modern experimental methods appear to offer great promise for unraveling the problems of nuclear binding relevant to the $r$-process. 
All the experimental progress to date represents a step forward in putting $r$-process simulations on solid ground. The present experimental methods also 
offer an excellent pathway for pursuing further mass measurements toward n-rich nuclei. 
It should be emphasized, however, that we are still far from collecting a complete nuclear mass database for $r$-process simulations. 
In fact, the experiments are just ``scratching the $r$-process path'' near the neutron magic numbers 50 and 82. We may summarize these points in a familiar way: More and
better data are needed.

High-resolution, high-precision, high-speed, and high-efficiency selective facilities for mass measurements of n-rich nuclei still need to be pursued in the coming years. Improved detection of heavier ions must be made to obtain good timing and position resolutions
  for the ions of interest. 
In addition to the need for improved techniques, $r$-process studies are limited mainly by the RIB intensity to the narrow isospin range of stable beams. The next generation RIB facilities like FAIR in Germany, FRIB in USA, now are under constructions, aiming at higher RIB intensity and purity.
One concept for overcoming this limit is to use  
the secondary fragmentation of neutron-rich nuclides (like $^{132}$Sn) produced by fission~\cite{Tanihata2008NIMB4067}. This idea has been discussed in EURISOL project.
   A new project, the China Advanced Rare Ion Beam Facility~\cite{BRIF-CARIF2011CS}
adopts this concept. It will be based on the China Advanced Research Reactor with ISOL of fission fragments, post-acceleration to 150 MeV/u, and fragmentation of a neutron-rich fission fragment beam such as $^{132}$Sn. This combination can offer an extremely neutron-rich beam, with an intensity that may be higher by one to two orders of magnitude than the current one. To perform mass measurements of fission fragments, MR-TOF-MS, which combines high resolution and compact size, 
will be an ideal choice. It can be located immediately after the ISOL and measure the masses of those ``easy but good'' exotic nuclei with long half-lives and high fission yields, such as $^{132}$Sn. More importantly, it can be allocated at the site immediately after in-flight separation to cover most extremely neutron-rich fragments. 


Nevertheless, the low yield of these nuclei is currently still a
serious challenge, and this problem will be faced even by next-generation radioactive facilities worldwide. 
Therefore, in the foreseeable future, we will still have to resort to nuclear theories for modeling the $r$-process, especially for $r$-nuclei at $N=126$. 
Unfortunately, even the most accurate mass models have rms mass errors of about 500 keV for known nuclei~\cite{Lunney2003Rev.Mod.Phys.1021,Adam2014PhysRevC.89.024311,RMF2014FronPhys529}, and this 
is clearly insufficient for a reliable $r$-process simulation. One of the most successful models
is the WS4 model~\cite{Wang2014PLB215}, whose rms deviation with respect to the available mass data is about 0.3 MeV.  This accuracy is generally comparable to that of the known experimental data for most exotic nuclei or to that of the TOF-B$\rho$-MS method. However, the problem here is that the reliability of computations of nuclei with unknown masses degrades when the computations extrapolate further away from the known mass surface~\cite{Sun2008Phys.Rev.C025806,Niu2012CPS55,Zhao2012Phys.Rev.C64324,Adam2014PhysRevC.90.017302}. 
For very neutron-rich nuclei, the predictions of various mass models can differ by more than a few mega-electron volts. 
Hybrid models including the use of the radial basis function approach~\cite{Niu2013PhysRevC.88.024325,Niu2014PhysRevC.90.014303},   
the residual proton--neutron interactions~\cite{Jiang2012PhysRevC.85.054303}, and the systematics of
the alpha decay energies~\cite{Li2013PhysRevC.88.057303} 
can provide highly precise predictions for nuclei near the known mass surface; however, the growth of the intrinsic error for nuclei far from the known region remains a serious problem.
In this sense, any reliable experimental data can serve as a critical test of existing  
mass models and guide their further development. 
  
Finally, we would like to draw attention to one of the major open questions for both $r$-process and nuclear structure studies: will the magic neutron numbers 50 and 82 remain for the 
nuclei along the $r$-process path? 
Assuming the validity of the single-particle picture and furthermore no rearrangement at the single-particle level when additional nucleons are added, 
one way to evaluate the shell closure is to refer to the effective neutron shell gap energies. They are defined as the difference in the two-neutron separation energies: 
$\Delta_n(Z, A)= S_{2n}(Z, A)-S_{2n}(Z, A+2)$. 
Experimental shell gap energies are plotted in Fig.~\ref{fig:nshell} for $N = 50$ and 82. 
Various nuclear models, i.e., the FRDM~\cite{Moller1997ADNDT}, WS4~\cite{Wang2014PLB215}, and HFB-21~\cite{HFB-21}, and 
a local mass relation, the so-called $\delta V_{np}$~\cite{Fu2011PhysRevC.84.034311,Jiang2012PhysRevC.85.054303}, are also shown for comparison. 
Clearly, a few more experimental points are needed to pin down the presence or absence of a shell quenching effect at $N=$50 and 82, as well as the accuracy of nuclear models.

\acknowledgments
This review is based entirely on the common efforts of our
colleagues working over many years in different
collaborations. This work was supported partially by the National Natural Science Foundation of China
(Nos. 10975008, 11035007, 11105010, 11235002, 11128510, 11475014),  
the Program for New Century Excellent Talents in University (No. NCET-09-0031), and Helmholtz-CAS Joint Research Group HC JRG-108..

\bibliography{Draft_20150803-final-pdf.bbl}
\end{document}